\documentclass{cernyrep}
\usepackage{texnames}
\usepackage[T1]{fontenc}

\usepackage{endnotes}       % use \theendnotes after bibliography
\usepackage{color}
\newcommand{\ced}[1]{\textcolor{black}{#1}}

\def\rmi{\mathrm{i}} % imaginary
\def\rmd{\mathrm{d}} % differential
\def\rme{\mathrm{e}} % exponential / electron
\def\u{\:}           % space in between number and unit, and between parts of units
\def\ee{\mathrm{ee}} % imaginary
\def\ii{\mathrm{ii}} % imaginary
\def\ei{\mathrm{ei}} % imaginary
\def\ie{\mathrm{ie}} % imaginary
\renewcommand{\vec}[1]{\mbox{\boldmath{$#1$}}} % changed vector overarrows to bold italic
\usepackage{amsfonts}
 \DeclareSymbolFont{upgreek}{U}{eur}{m}{n}
 \DeclareMathSymbol{\rmalpha}{0}{upgreek}{"0B}
 \DeclareMathSymbol{\rmbeta}{0}{upgreek}{"0C}
 \DeclareMathSymbol{\rmgamma}{0}{upgreek}{"0D}
 \DeclareMathSymbol{\rmdelta}{0}{upgreek}{"0E}
 \DeclareMathSymbol{\rmmu}{0}{upgreek}{"16}
 \DeclareMathSymbol{\rmpi}{0}{upgreek}{"19}
 \DeclareMathSymbol{\rmsigma}{0}{upgreek}{"1B}
 \DeclareMathSymbol{\rmphi}{0}{upgreek}{"1E}
 \DeclareMathSymbol{\rmomega}{0}{upgreek}{"21}
\begin{document}
\title{Production of High-Intensity, Highly Charged Ions}

\author{S. Gammino}

\institute{Istituto Nazionale di Fisica Nucleare, Laboratori Nazionali del Sud, Catania, Italy}

\maketitle % this produces the title block

\begin{abstract}
In the past three decades, the development of nuclear physics facilities for fundamental and applied science purposes has required an increasing current of multicharged ion beams. Multiple ionization implies the formation of dense and energetic plasmas, which, in turn, requires specific plasma trapping configurations. Two types of ion source have been able to produce very high charge states in a reliable and reproducible way: electron beam ion sources (EBIS) and electron cyclotron resonance ion sources (ECRIS). Multiple ionization is also obtained in laser-generated plasmas (laser ion sources (LIS)), where the high-energy electrons and the extremely high electron density allow step-by-step ionization, but the reproducibility is poor. This chapter discusses the atomic physics background at the basis of the production of highly charged ions and describes the scientific and technological features of the most advanced ion sources. Particular attention is paid to ECRIS and the latest developments, since they now represent the most effective and reliable machines for modern accelerators.
\end{abstract}

\section{Introduction}

Research on ion sources is basically \ced{dictated} by the growing \ced{need for} intense currents of multicharged ions \ced{in} modern nuclear physics facilities. Ion acceleration in a cyclotron obeys the scaling law \mbox{$E \propto {q}/{A}^2$}, where $q$ is the charge state and $A$ is the mass number, so the advantages of highly charged ion (HCI) sources is evident. Intense currents are \ced{required} in order to reduce the acquisition time in low-cross-section nuclear physics experiments, or to increase the production rate of exotic beams in modern facilities such as SPIRAL2, FAIR and so on. \ced{To quote one} example, the FRIB (Facility for Rare Isotope Beams) project at MSU, Michigan, USA, requires $I \geq 280{\u}\rmmu$A for U$^{34+}$. Similarly, the FAIR facility at GSI, \ced{Darmstadt, Germany,} will be based on a 1{\u}mA beam of U$^{28+}$. Multicharged ion production can be obtained in the presence of  a flow of electrons or in a plasma. \ced{The various} ionization techniques differ \ced{from each other in terms of} the electron temperature, electron density and plasma lifetime that can be achieved. The simplest way to create plasmas consists of electrical discharges in  vacuum tubes; other ways use electron beams passing through neutral gases or electromagnetic waves interacting with gases or vapours in the presence of a well-shaped magnetic field. The latter method is preferable for stable and reproducible HCI formation, since the magnetic field \ced{lengthens} the time spent by the ions in the volume where multiple ionizing collisions with energetic electrons occur.
When the plasma heating is provided by means of microwaves and \ced{under} the electron cyclotron resonance (ECR) condition, the plasmas are named ECR plasmas.

The characterization of the different types of ion sources can be done by taking into account some fundamental quantities (\textit{quality factor}). In particular, as  ionization to high charge states requires long times, it is clear that the ion confinement time will be a crucial parameter to determine the maximum achievable charge state. The number of available electrons  is also important, because it is clearly related to the number of electrons able to ionize the atoms down to inner shells. Hence a qualitative criterion for evaluating the ability of a given plasma to provide HCIs is given by the quality parameter $Q$, defined as the product of the electron density $n_\rme$ with the ion confinement time $\tau_\rmi$:
\begin{equation}
\label{fattoreQ}
Q=n_\rme \tau _i .
\end{equation}
The following proportionality is valid:
\begin{equation}
\label{eq,qmed_propto_Q}
\langle q \rangle \propto n_\rme \tau_\rmi .
\end{equation}

Motivations leading to such rules will \ced{become clearer in the following sections}. \ced{Here, it is worth mentioning} that each source, according to its peculiar characteristics, will have a specific capability to increase the electron density or the ion confinement time, \ced{and thus} to maximize the $Q$ parameter.
Although $Q$ can be considered as the main parameter to determine the quality of ion source performance, the electron temperature $T_\rme$ (or more generally the average electron temperature in the plasma) also plays a key role: it fixes the maximum charge state that can be achieved depending on the given ionization potential.
In terms of extracted currents, a confinement time \ced{that is too long} would limit the amount of ions extracted from the system per \ced{unit time}. Well-confined plasmas can produce extremely high charge states, but current intensity will be modest, i.e.
\begin{equation}
\label{eq,current_vs_ne_taui}
I \propto \frac{n_\rme}{\tau_\rmi} .
\end{equation}
Since Eq.~(\ref{eq,current_vs_ne_taui}) displays an inverse proportionality between current and confinement time, and a direct dependence on the electron density (as in the case of $Q$), the electron density $n_\rme$ becomes the key parameter if one wants to maximize both currents and charge states.

In order to complete the set of quality parameters, we have to take into account the total background pressure in the chamber where the plasma is created. Fully ionized plasmas are rare, and in many cases the neutral atoms and the lowly charged ions may decrease the mean charge state because of charge exchange processes (whose cross-section is considerably higher than the cross-section of electron recombination). Hence another quality criterion can be inferred, and it states that
\begin{equation}
\label{eq,chargeexchange}
\frac{n_0}{n_\rme} \leq f\left(T_\rme,A,z\right) ,
\end{equation}
where $f$ is a function of the atomic mass $A$, the atomic number $z$ and the electron temperature $T_\rme$.
Finally, in order to perfectly match the ion sources with the accelerators, the beam quality must fit the accelerator constraints, in terms of brightness and emittance.

Additional characteristics of ion sources are envisaged, \ced{such as} the reliability, the lifetime of the system, the maintenance and, last but not least, the cost (for building and for ordinary operations).

\section{Atomic physics background for HCI production}

\ced{The dynamics of the collisions} among the plasma particles \ced{is} of primary importance for ionization rate determination and for dealing with a great number of other phenomena such as electron and ion lifetime evaluation in magnetically confined plasmas. An atomic physics background is therefore needed to \ced{distinguish between} the most significant plasma parameters influencing ion source performance. Basically, the collision dynamics \ced{is} mainly determined by the electron temperature, the electron density and the background pressure.

It is convenient to introduce some physical quantities that are useful for the treatment of collisional dynamics. In particular, once \ced{we have} defined the \textit{cross-section}, we can introduce the mean free path, the mean time between a collision and the following one, and finally the collision frequency. The analysis of the various collisions in plasmas will be carried out principally in terms of the collision frequency.

If $\sigma$ is the cross-section of a given collision, then the mean free path will be given by
\begin{equation}
\label{liberocammino}
\lambda_{\rm m} = \frac{1}{n \sigma} ,
\end{equation}
where $n$ is the density.\footnote{Here we \ced{denote the} general density by $n$. The electron and ion densities are usually distinguished by using the notation $n_\rme$ and $n_i$.} Once the $\lambda_{\rm m}$ parameter \ced{has been defined}, the time between two successive collisions will be
\begin{equation}
\label{taugen}
\tau = \frac{\lambda_{\rm m}}{v} ,
\end{equation}
where $v$ is the velocity of the colliding particles. The collision frequency can be easily determined from \eqref{taugen}, and is given by
\begin{equation}
\label{freqcollgen}
\nu=\frac{1}{\tau}=\frac{v}{\lambda_{\rm m}}=n\sigma v .
\end{equation}
Generally, in plasmas, all the particles (electrons, ions and neutrals) follow some energy distribution function,\footnote{In many cases it is convenient to use a Maxwellian-like distribution, for which a temperature can be determined. \ced{However}, for low-pressure and high-temperature plasmas, the energy distribution function does not follow a pure Maxwellian shape, but it can nevertheless be viewed as a superposition of several plasma populations with Maxwellian distributions at different temperatures.} and so we cannot consider a single velocity in \eqref{freqcollgen}. This expression can be still used to determine $\nu$, but we have to average\footnote{We must average not only over $v$, but also over $\sigma$ because in general \ced{we have} $\sigma\equiv\sigma(v)$.} over all the possible values of $\sigma v$. Then we obtain
\begin{equation}
\label{freqMax}
\nu=n\overline{\sigma v}.
\end{equation}

\ced{The assumptions made so far} are quite general and \ced{nothing specific has been said} about the several types of plasma collisions.
A schematic overview of the main plasma collision processes, and a classification, have been given according to the number of colliding particles: \textit{multiple collisions} are characteristic of plasmas, because they occur thanks to the \textit{long-range Coulomb interaction}; and \textit{binary collisions} are more similar to the collisions in gaseous systems. However, in this case, non-elastic and ionizing collisions occur because of the high energy content of the plasma particles, especially of electrons.

\subsection{Multiple-scattering collisions}

Note that in plasmas any electric field can be felt by charged particles only within a distance that is named the \textit{Debye length},
\begin{equation}
\label{lunghezzaDebye}
    \Lambda_{\rm D}=\left(\frac{\epsilon_0 K_{\rm B} T_\rme}{n
    e^2}\right)^{{1}/{2}}.
\end{equation}
This means that a proof particle feels the electromagnetic interaction due only to those plasma particles located inside the Debye sphere.\footnote{The Debye sphere is the sphere with radius $\Lambda_{\rm D}$.}
\ced{However, the number of particles included in the Debye sphere is still quite large (of the order of several thousands). A test particle, therefore, will be subjected to multiple interactions for which is more convenient to evaluate the cross section for a cumulative deflection of 90$^\circ$.} Then we can  define a cross-section, $\sigma_{90^\circ}$, given by\footnote{The quantities in the formula are usually expressed in the c.g.s.\ system}
\begin{equation}
\label{eq,sigma_novanta}
\sigma_{90^\circ} = 8\pi \left(\frac{z_1 z_2 e^2}{M v^2}\right)^2
    \ln\left(\frac{\Lambda_{\rm D}}{b_{\rm min}}\right) ,
\end{equation}
where $z_1$ and $z_2$ are the charges of the two particles, $e$ is the electron charge, $M$ is the mass of the colliding particle (if the other particles are much heavier) and the term ${\Lambda_{\rm D}}/{b_{\rm min}}$ is the so-called Coulomb logarithm. This equation, except for the logarithm, is practically the Rutherford formula for charged particle scattering (binary collision). The Rutherford cross-section is almost two orders of magnitude lower than \eqref{eq,sigma_novanta}. This means that a 90$^\circ$ scattering due to a single collision is much less probable than multiple deflections due to many collisions.\footnote{This result was formerly guessed by Chandrasekhar \cite{ch}.}

A more precise representation of the multiple collisions can be given in the centre-of-mass system, and the collision frequencies can be calculated by taking into account the collisions among the two different plasma species: electrons and ions. Omitting, for the sake of brevity, the mathematical description, we can use the final formulas to understand the role that each type of collision plays in the plasma:
\begin{eqnarray}
\label{ristretta2}
\nu_{90^\circ}^{\ee}&=&\frac{1}{\tau_{\rm sp}} = 5\times10^{-6}n
\frac{\ln({\Lambda_{\rm D}}/{b})}{T_\rme^{{3}/{2}}} , \\
\nu_{90^\circ}^{\ei}&=&\frac{1}{\tau_{\rm sp}} \sim 2\times10^{-6}zn
\frac{\ln({\Lambda_{\rm D}}/{b})}{T_\rme^{{3}/{2}}} , \\
\nu_{90^\circ}^{\ii}&=&\frac{1}{\tau_{\rm sp}} \sim z^4 \left(\frac{m_\rme}{m_\rmi}\right)^{{1}/{2}}\left(\frac{T_\rme}{T_\rmi}\right)^{{3}/{2}}\nu^{\ee}_{90^\circ},
\end{eqnarray}
where $n_\rme$ is in cm$^{-3}$ and $T_\rme$ is in eV.

Each frequency is usually called the \textit{characteristic Spitzer collision frequency}, while $\tau_{sp}$ is the characteristic Spitzer collision time elapsing on average between two subsequent collisions. Once the collision frequency\footnote{We talk about the collision frequency for the sake of simplicity, but strictly speaking the term `collision' is incorrect, as we should speak about 90$^\circ$ scattering.} is given, we can easily determine the time between two successive events: $\nu_{90^\circ}^{\ee}\simeq\nu_{90^\circ}^{\ei}$ and $\tau_{90^\circ}^{\ee}\simeq\tau_{90^\circ}^{\ei}$. \ced{In contrast}, if we consider ions colliding with electrons (ion--electron collisions), we have
\begin{equation}
\tau_{90^\circ}^{\ie}\simeq\left(\frac{m_\rmi}{m_\rme}\right)\tau_{90^\circ}^{\ei}.
\end{equation}
Thermalization among different plasma species is governed by collisions. The time needed for a $90^\circ$ deflection and that needed for energy exchange are defined as
\begin{eqnarray}
\label{relencoll}
\tau_{\rm m}^{\ee}&\sim&\tau_{90^\circ}^{\ee}\sim\tau_{90^\circ}^{\ei},\\
\tau_{\rm m}^{\ii}&\sim&\tau_{90^\circ}^{\ii}
                   \sim\left(\frac{m_\rmi}{m_\rme}\right)^{{1}/{2}}\tau_{90^\circ}^{\ei},\\
\tau_{\rm m}^{\ei}&\sim&\tau_m^{\ie}\sim\frac{m_\rmi}{m_\rme}\tau_{90^\circ}^{\ei}.
\end{eqnarray}
The subscript "m" indicates the average time needed for energy sharing collisions.

The Spitzer collisions are the dominant processes in highly ionized and low-pressure plasmas, and totally regulate the thermalization processes. From the above formulas, it \ced{can be seen that} $\tau_{\rm m}^{\ee}\sim\tau_{90^\circ}^{\ee}$ and $\tau_{\rm m}^{\ei} \gg\tau_{90^\circ}^{\ei}$, implying a much easier electron--electron thermalization than electron--ion one. As an example, at typical electron energies ranging from a few electronvolts\footnote{This temperature is typical for the ion sources used to produce lowly charged ion beams at high currents (tens of mA): these sources are usually defined as microwave discharge ion sources.} to some kiloelectronvolts,\footnote{For ECRIS plasmas for highly charged ions.} the value of $\nu_{\ee}$ varies from $10^5\mbox{--}10^6${\u}s$^{-1}$ ($T_\rme=10${\u}eV) to $10^{2}\mbox{--}10^{3}${\u}s$^{-1}$ ($T_\rme=1${\u}keV), so that the electron thermalization requires \ced{a few microseconds}, in the first case, or \ced{milliseconds} in the second case. The ion thermalization proceeds on a ${m_\rmi}/{m_\rme}$ (i.e.\ about $10^3\mbox{--}10^4$) longer timescale. In an electron cyclotron resonance ion source (ECRIS), where the ion lifetime is shorter than the time required for the ion collisional heating, the ions remain cold. This result has an immediate consequence on the emittance, which can thereby be maintained sufficiently low.\footnote{It strongly depends on the ion temperature.}
Another consequence deriving from the set of equations (\ref{relencoll}) is that all the electrostatic collision frequencies are much smaller than some other characteristic frequency of the plasma, in particular the Larmor frequency (if the plasma is magnetized) and the microwave frequency (if the plasma is sustained by electromagnetic fields):
\begin{equation}
\omega_{\rm RF}\gg\nu_{\ee}\quad\mbox{and also}\quad\omega_{\rm RF}\gg\nu_{\ei}.
\end{equation}
Under this condition the plasma is called \textit{collisionless}.

\subsection{Binary collisions}

The binary collisions in plasmas can be divided into two groups: the  \textit{electron--neutral collisions} (elastic and inelastic) and the  \textit{ionizing collisions}. The former play an important role especially in the case of high-pressure plasmas with a low degree of ionization and a low electron temperature.\footnote{By high pressure, we mean, in this case, $10^{-3}\mbox{--}10^{-4}${\u}mbar, i.e.\ the operating pressures of a microwave discharge plasma (MDP). Low temperature here means $T_\rme \leq 15${\u}eV, which also in this case is typical of an MDP.} These collisions provide the thermalization in this kind of plasma, playing the same role as the Spitzer collisions in the case of highly ionized and low-pressure plasmas.
As for the creation of multicharged ions in ECRIS, these collisions are not so important, and so we focus our attention on the ionizing collisions.

\begin{figure}[htbp]
\begin{center}
\includegraphics[width=0.85\textwidth]{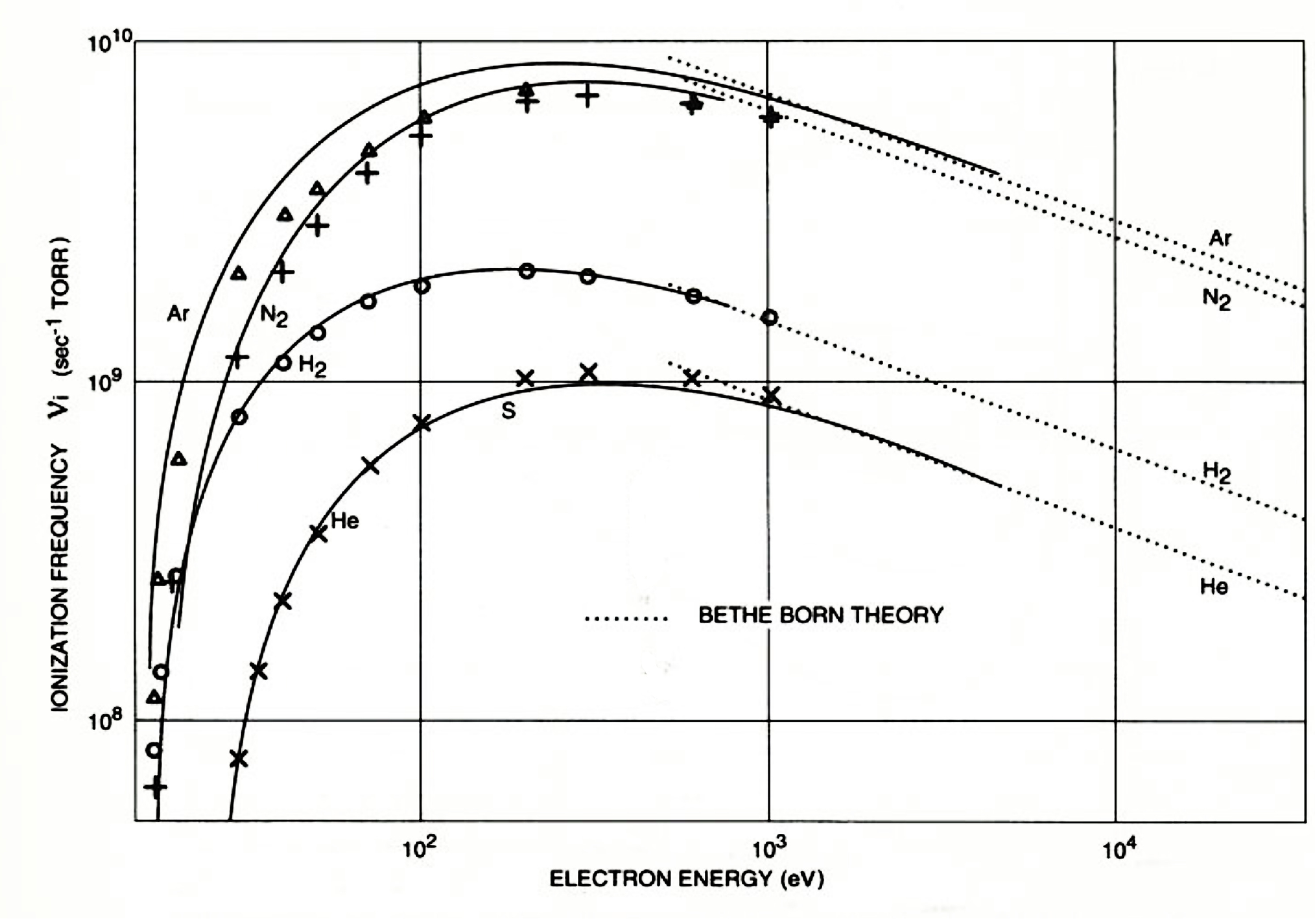}
\end{center}
\caption{Ionization frequencies versus the electron energy. For high electron energies the experimental data have been fitted with the Bethe--Born theoretical formula \cite{gell}. \label{fig,ionization_frequency}}
\end{figure}

Figure \ref{fig,ionization_frequency} shows the trend of the ionization frequency versus the electron energy for different ion species. It can be seen that $\nu_\rmi$ is parametrized by the background pressure. Hence, to determine the real ionization rate, we have to multiply the frequency shown in the figure by the operating pressure of our ion source. In addition, the trend shown in Fig.~\ref{fig,ionization_frequency} considers all the possible ionizing events, i.e.\ the production rates for all the charge states that can be produced by a single collision. However, it can be demonstrated that generally a single ionizing collision expels just one electron from the atomic shells, thus requiring many collisions to achieve highly charged ions. For higher electron energies, $\nu_\rmi$ can be approximated by the Bethe--Born formula:
\begin{equation}
q_\rmi \propto \frac{1}{E} \ln E \quad \mbox{or} \quad
\nu_\rmi\propto\frac{1}{\sqrt{E}}\ln E.
\end{equation}
It is clear from Fig.~\ref{fig,ionization_frequency} that the ionizing frequency increases after a threshold value (corresponding to the ionization potential), and then slowly decreases for higher electron energies. If we take into account the usual Maxwellian distribution function of the electrons, we may replace the electron energy on the $x$ axis of Fig.~\ref{fig,ionization_frequency} with the electron temperature.  For example, in a microwave discharge plasma the electron temperature is usually $\sim10${\u}eV, but the atoms can be easily ionized because of the distribution tail electron component, whose energy is high enough to provide effective ionizing collisions.

In the case of ECRIS, where the production of highly charged ions is of primary importance, a particular relationship among electron density, electron temperature and ion confinement time has to be achieved. This discussion leads to the so-called  \textit{multicharged ion (MI) production criterion}. As mentioned above, the ionization down to inner shells of atoms requires a  \textit{step-by-step} process, because the cross-section for single ionization per collision is much higher than \ced{that for} multi-ionization.

\begin{figure}[htbp]
\begin{center}
\includegraphics[width=0.45\textwidth]{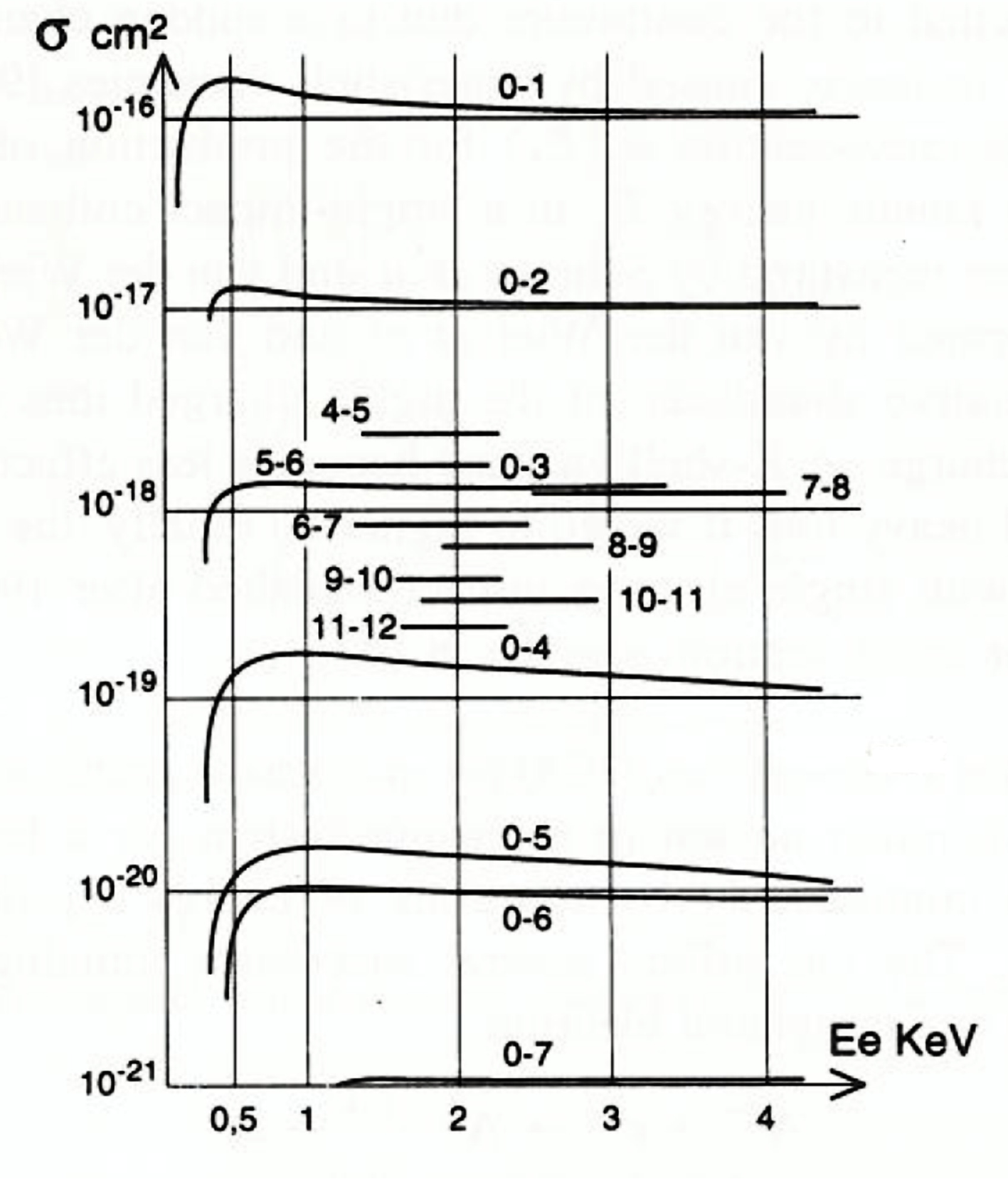}
\end{center}
\caption{Single-impact ionization cross-section $\sigma_{0_z}$ compared to successive-impact ionization cross-sections versus electron energy (Ar). \label{fig,cross_section_MI} }
\end{figure}

This is confirmed by Fig.~\ref{fig,cross_section_MI}, which shows that the cross-section for single ionization due to a single collision is of the order of $10^{-16}${\u}cm$^2$, and then, for multi-ionizations, it rapidly decreases. It \ced{turns} out that the probability to expel two electrons is one order of magnitude lower; to expel three electrons it is two orders of magnitude lower; and so on. The figure also shows the cross-section for single ionizations, but in the case of highly charged ions (transitions $z \rightarrow z+1$).

The time needed for the transition from charge state $z_1$ to charge state $z_2$ takes a time, on average, of
 \begin{equation}
\tau_{z_1z_2}=\frac{1}{[n_\rme\sigma_{z_1z_2}(v_\rme)v_\rme]},
\end{equation}
and averaging over a Maxwellian distribution\footnote{We usually average the collision parameters over a Maxwellian distribution. The energy distributions of ECRIS plasmas differ from a Maxwellian trend, but they can be considered as superpositions of three different distributions, each one having its own temperature. Then, for the sake of simplicity, we assume that the results obtained for a Maxwellian distribution are valid for each plasma population, with the further consideration that the warm component is the most effective to produce highly charged ions. \ced{From now on}, we will use the parameters of the `warm' ($\sim$\,keV) population to determine the ionizing frequencies of ECRIS plasmas.} we have
\begin{eqnarray}
\label{eq,time_of_ionization}
\tau_{z}&=&\frac{1}{[n_\rme S_{z}(T_\rme)]},\\
S_{z}(T_\rme)&=&\langle\sigma_{z}(v_\rme)v_\rme\rangle(T_\rme).
\end{eqnarray}
If the ion confinement time is longer than the time required for the given ionization (i.e.\ the time defined in Eq.~\eqref{eq,time_of_ionization}), the transition $z\rightarrow z+1$ takes place. Then the condition to produce highly charged ions in plasmas can be written as
\begin{equation}
\label{criterioMI}
n_\rme\tau_\rmi\geq\frac{1}{[S_{z}(T_\rme)]}.
\end{equation}

In plasmas with long confinement times, the step-by-step ionization provides highly charged ions.
Equation \eqref{criterioMI} can be rewritten by substituting the $S_{z}(T_\rme)$ parameter, obtaining
\begin{equation}
\label{eq,MIcriterion}
\xi n_\rme \tau_\rmi
\geq 5 \times 10^{4} (T_{\rme}^{\rm opt})^{{3}/{2}},
\end{equation}
where $\xi\equiv\sum_{j=1}^{N}q_j$ is the number of subshells in the atomic outer shells.

\begin{figure}[htbp]
\begin{center}
\includegraphics[width=0.5\textwidth]{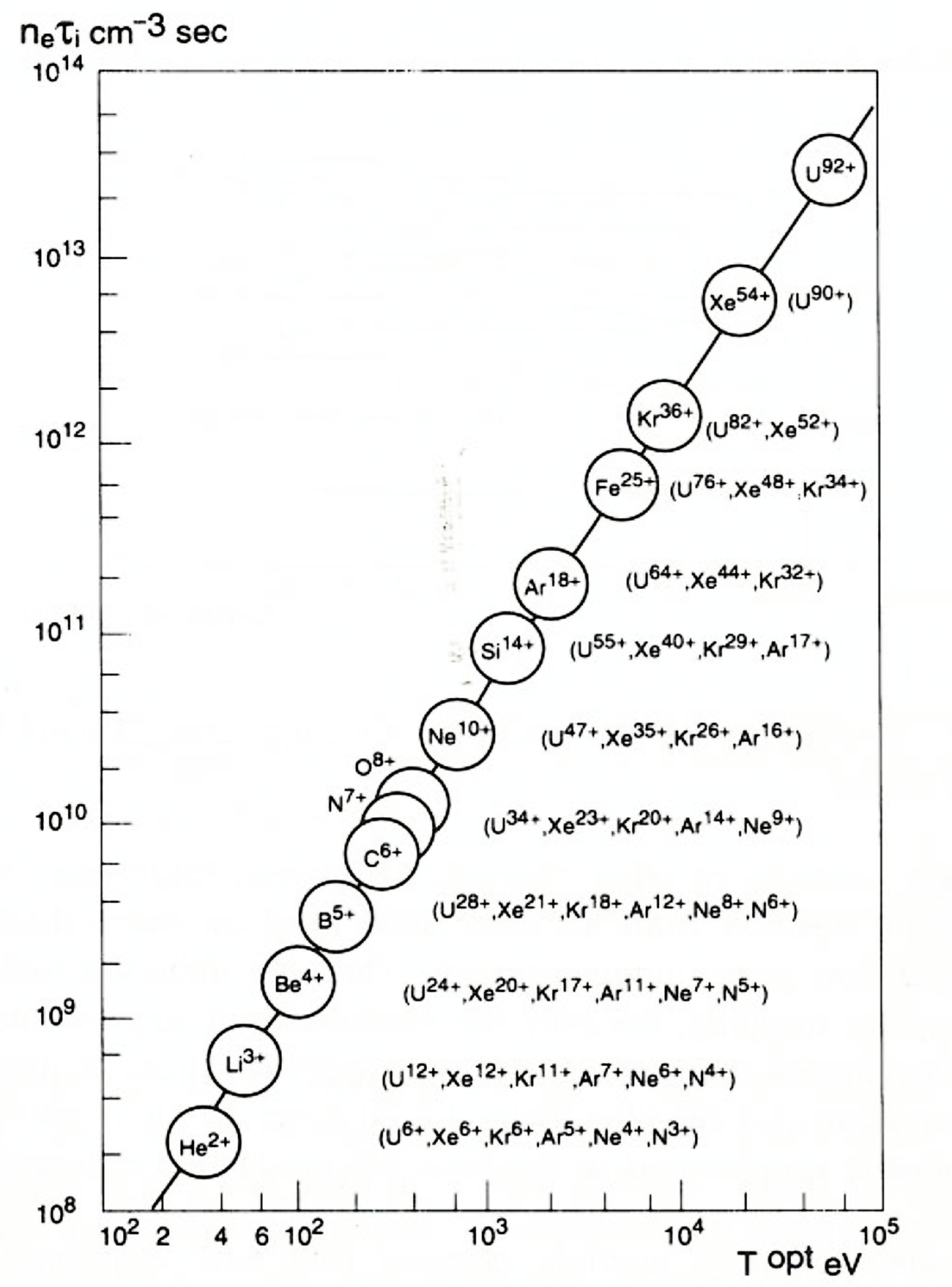}
\end{center}
\caption{Golovanivsky's plot of the $(n_\rme\tau_\rmi)$ criteria for the production of highly charged ions. The ions enclosed by circles are completely stripped; then some combinations of electron temperature, electron density and ion confinement time allow one to produce completely stripped ions. Inside the brackets, incompletely stripped ions are shown: they can be produced with the corresponding plasma parameter of ions enclosed in circles. \label{Golovanivsky_Plot} }
\end{figure}

The criterion that comes from Eq.~\eqref{eq,MIcriterion} is of paramount importance, as it fixes the plasma operational parameters needed for good-performance HCI sources. Some numerical examples can be given \ced{for} the required densities, confinement times and electron temperatures. If one aims to obtain completely stripped light ions,  then $T_\rme\simeq 5${\u}eV, with $n_\rme\tau_\rmi \geq 10^{10}${\u}cm$^{-3}$, is needed. The required electron temperature is given by the energy threshold of the given charge state, to obtain:
\begin{equation}
T_\rme^{\rm opt}\simeq 5W_{\rm thr}.
\end{equation}
A more general picture of the criteria linked to the plasma parameters is given by Fig.~\ref{Golovanivsky_Plot}, which features what combinations of temperatures, densities and confinement times are able to produce different ion species with different charge states. This is called a \textit{Golovanivsky plot}, as it was presented for the first time by the late Russian scientist more than 20 years ago.

\subsection{Recombination effects}

The charge state of highly charged ions may decrease because of some electron--ion recombination process. The recombination mechanisms are of three different types:
\begin{itemize}
\item  \textit{Radiative recombination.} Free electrons are captured by ions with the \ced{subsequent} emission of a photon.
\item  \textit{Dielectronic recombination.} This is a three-step process, where an electron first excites an ion, then it is captured, and finally a photon is emitted.
\item  \textit{Charge exchange recombination.} Highly charged ions, colliding with low charge species and/or with neutrals, capture some electrons from the electronic cloud of the other particle, thus decreasing their own charge state.
\end{itemize}
By analysing the cross-sections of each mechanism, it \ced{turns} out that only charge exchange recombination plays a \ced{marked} role in ion sources. Because of its nature, the recombination rates must depend on the background pressure: the higher the number of neutral or low-charge-state ions, the higher is the probability of charge exchange collisions. Again, in order to calculate the correct parameters for ECRIS, the time needed for a single charge exchange process must be evaluated. The charge exchange collision time must be longer than the ionization time for a given charge state (see Eq.~\eqref{eq,MIcriterion}). Bypassing the mathematical passages, we can finally write
\begin{equation}
\label{qualityNeutri}
\frac{n_0}{n_\rme}\leq
7\times10^3\xi\frac{\sqrt{A}}{z[T_\rme^{\rm opt}]^{{3}/{2}}},
\end{equation}
where $n_0$ is the density of neutrals, and $A$ is the atomic mass number.

Then, in order to obtain \ced{fully stripped argon ions} (Ar$^{18+}$) at $n_\rme\simeq 10^{12}${\u}cm$^{-3}$, we need an operational pressure of $\sim10^{-7}${\u}mbar, corresponding to a number of neutrals per cubic centimetre $\sim10^9${\u}cm$^{-3}$. These numbers are typical of many modern ECRIS, where the minimum operational pressure is between 10$^{-8}$ and $5\times10^{-7}${\u}mbar.

In summary, the analysis of collision dynamics in plasmas \ced{highlights} the importance of low background pressure and of plasma confinement by means of magnetic fields. For low pressures ($p<10^{-4}${\u}mbar), the mean free path for ionizing collisions becomes \ced{longer} than the plasma itself:\footnote{Plasma dimensions are comparable with ones of the metallic chambers usually employed in ion sources for sustaining the plasma, i.e.\ some tens of centimetres.} $\lambda_\rmi\geq300${\u}cm.
Only by means of a magnetic field is it possible to have some collisions in times shorter than the ion lifetime. Electrons spiral around the field lines and in addition they are reflected by a magnetic mirror, so that they go back and forth inside the magnetic trap and collide with neutral atoms or charged ions.

\section{Ion sources for high charge states and high current intensity}
\label{par,sorgentivarie}

\ced{The following are the} principal ion sources able \ced{to fulfil the needs of} modern accelerators in terms of high charge states and high current:
\begin{enumerate}
    \item  Electron beam ion sources (EBIS)
    \item  Laser ion sources (LIS)
    \item  Electron cyclotron ion sources (ECRIS)
\end{enumerate}
On the basis of the discussion in the previous section, a comparison of the data (both for currents and for charge states) \ced{often favours} ECRIS, which in fact are the most used in laboratories all over the world. Otherwise, for specific applications, EBIS offer remarkable advantages (beam quality), whereas LIS are able to provide very high current with a short pulse.

\subsection{Electron beam ion sources (EBIS)}
\label{sub,sorgentiEBIS}

The EBIS sources \cite{wolf} follow a simple principle: an electron beam is focused by means of a strong magnetic field generated by a solenoid, thus creating a plasma with ions confined by the potential well due to the focused electron beam. A simplified picture of this mechanism is shown in Fig.~\ref{fig,EBIS}. The atoms filling the plasma chamber are ionized by the electron beams. Once they became ions, they are confined by the potential well, and their confinement time increases as their charge state increases.

Cylindrical electrodes surrounding the device can be used to increase the potential well, thus increasing the ion confinement time and producing higher charge state.
Figure \ref{fig,EBIS} schematically describes an EBIS. The solenoids for plasma confinement are shown, and also the extraction system.

\begin{figure}[htpb]
\centering
\includegraphics[width=0.75\textwidth]{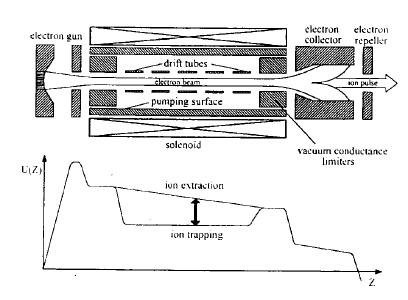}
\caption{EBIS scheme principles  \cite{wolf}\label{fig,EBIS}}
\end{figure}

EBIS are \ced{widely} used as the injectors for synchrotrons of completely stripped light ions (e.g.\ Dubna, Saclay or Stockholm). However, they can be also used as the injector of highly charged heavy ions (e.g. CERN or LNBL).

For very high charge states, the maximum currents are no higher than some nanoamps, but nevertheless they are able to produce charge states up to \ced{Xe$^{53+}$ and Xe$^{54+}$} (JINR Dubna, Russia).
Some experimental tests carried out at the Los Alamos National Laboratories, USA, have demonstrated that the creation of totally stripped uranium ions is possible, even if the extracted current is of the order of a few tens of ions per second. Although these machines are completely useless for accelerators, they allow one to carry out very interesting experiments for atomic physics.

In these devices, to reach such high charge states, the second term of the expression \eqref{eq,qmed_propto_Q} has been maximized. Because of this trapping optimization, these sources are usually named  \textit{electron beam ion traps} (EBIT). The expression \eqref{eq,current_vs_ne_taui} explains how only a few ions per second are produced in the Los Alamos machine, as the considerable increase of the confinement time is not accompanied by a corresponding increase of the electron density.

Finally, we can summarize the main advantages and disadvantages of EBIS:
 \begin{itemize}
    \item  \textit{Advantages}\\
    High charge states can be easily produced\\
    Beam quality is quite good because of the low-temperature ions (emittance is negatively affected by high-temperature ions)\\
    High possibility of obtaining various  pulse lengths or continuous-wave beams

    \item  \textit{Disadvantages}\\
    Low currents\\
    High realization costs\\
    Very complex device
\end{itemize}

\subsection{Laser ion sources (LIS)}
\label{sub,LIS}

Laser ion sources \cite{wolf} are based on   laser beams  focused onto a solid target surface  (Fig.~\ref{fig,LIS}).
If the power density of the laser beam is high enough, then a rapid vaporization of the target material occurs, and a  \textit{plasma plume} is generated, expanding at supersonic velocities (higher than $10^4${\u}m{\u}s$^{-1}$). Figure \ref{fig,plume} shows the expanding plume; this photo was \ced{taken} during a recent experiment carried out at LNS \cite{lns} with a Langmuir probe, able to characterize the plasma plume in terms of temperature and density for different times from a laser pulse. It helps to understand the plasma dimensions: the probe was \ced{placed} 1.5{\u}cm from the tantalum target, and the plume front reaches the metallic tip after approximately 1{\u}$\rmmu$s. The transverse plasma dimension is comparable with the probe tip length, about 1{\u}cm.

\begin{figure}[htbp]
\begin{center}
\includegraphics[width=0.75\textwidth]{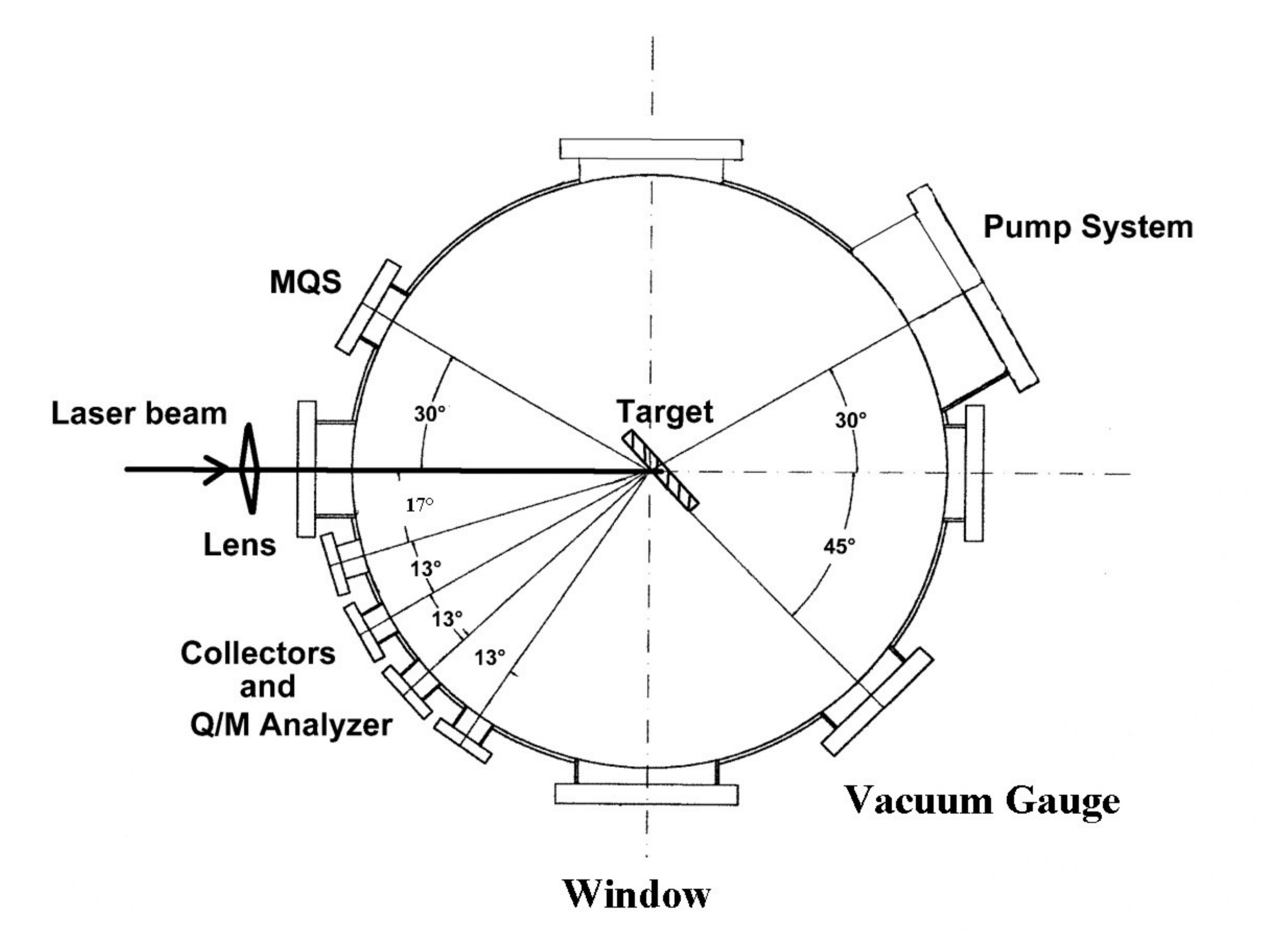}
\end{center}
\caption{\label {fig,LIS} Schematic view of a LIS source}
\end{figure}

\begin{figure}[htpb]
\centering
\includegraphics[width=0.49\textwidth]{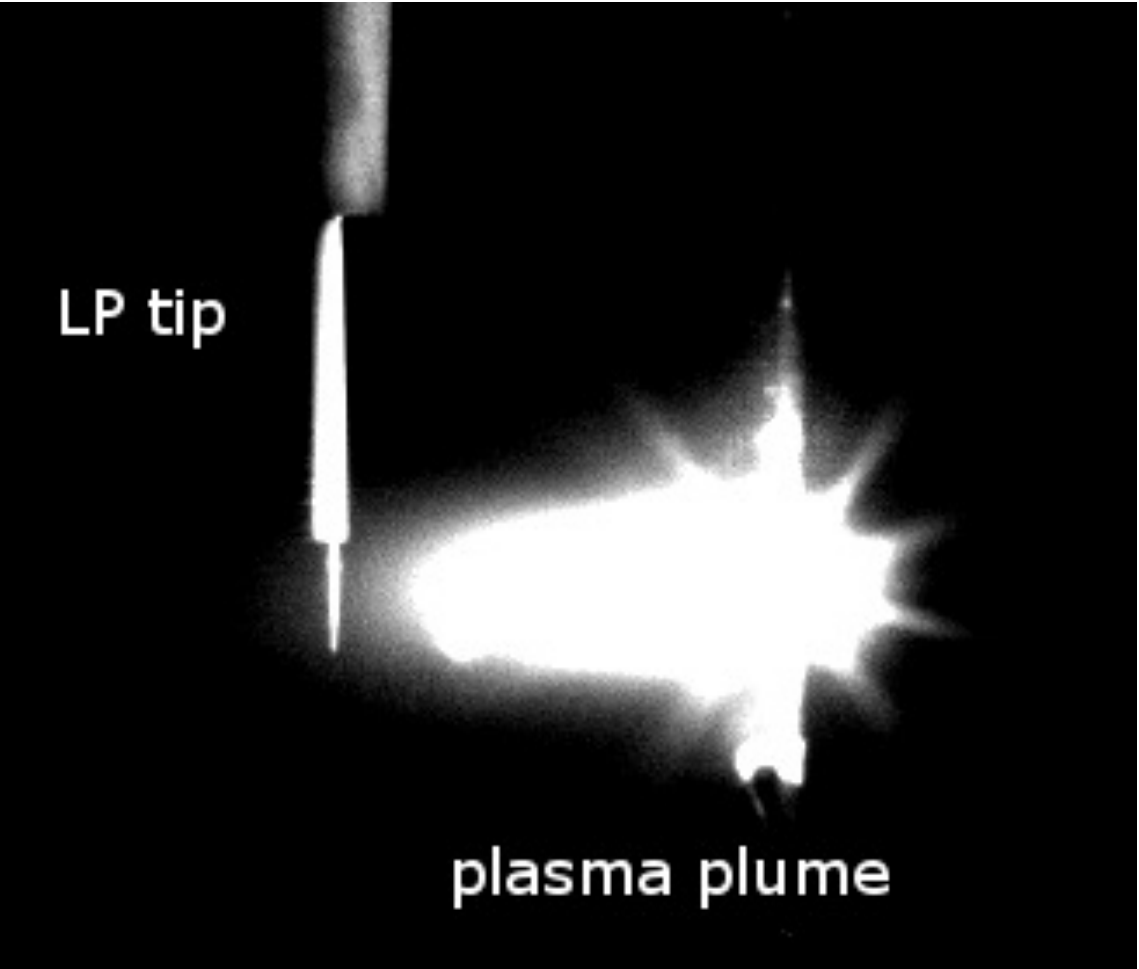}
\vspace{3mm}
\includegraphics[width=0.49\textwidth]{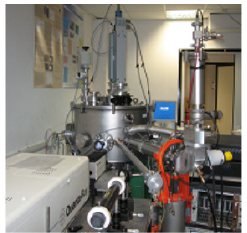}
\caption{\label{fig,plume} Plasma plume in a LIS source and the experimental apparatus for Langmuir probe measurements of laser-generated plasma properties.}
\end{figure}

Electrons generated during the target evaporation are heated by the remaining part of the laser radiation up to several tens of electronvolts when a Q-switched Nd:YAG laser, operating at 1064{\u}nm wavelength, with 9{\u}ns pulse duration and 1--900{\u}mJ energy, is used.
Such a laser operates at INFN-LNS, where a LIS is used instead of an ion source to study the physics of plasma produced by laser ablation.

Usually the laser beam is focused through a convergent lens onto a metal target (Ta, Ag, Cu, etc.) placed inside a vacuum chamber at $10^{-6}${\u}mbar. The laser beam incidence angle is lower than 90$^\circ$ with respect to the target surface and the spot diameter is 1{\u}mm. The plasma plume is emitted along the normal to the irradiated target whatever the direction of laser beam is with respect to the normal direction.

In general, the laser energy varies, from 1{\u}J to 50{\u}J, with  \textit{pulse lengths} up to 1{\u}$\rmmu$s and  \textit{repetition
rate} up to 1{\u}Hz for CO$_2$ lasers.
The electron temperature, $T_\rme$, and the charge state distribution depend on laser properties as
\begin{equation}
    T_\rme\approx(P\bar{Z})^{{2}/{7}},
\end{equation}
where $P$ is the laser power.
Figure \ref{fig,LIS} shows in particular the scheme of the LNS's laser source. The system is composed of different parts:
\begin{itemize}
\item laser beam generator;
\item optical system for laser focalization;
\item vacuum chamber where the target is located;
\item several diagnostics tools for analysis of ion properties (time-of-flight (TOF) measurements, ion energy analysers (IEA), etc.);
\item pumping system for vacuum.
\end{itemize}

The main limitation to the use of LIS in accelerator facilities is linked to the low repetition rate. This is mainly due to the laser pumping, which requires a lot of time, especially in the case of high-power, high-intensity lasers.
Anyway, as mentioned above, LIS are powerful methods to investigate non-equilibrium plasma physics, and they find applications in many fields of fundamental and applied physics (one of the most recent applications concerns cultural heritage).

The most important application of laser sources to accelerator physics consists of using the expanding plumes  as accelerators themselves. It has been demonstrated that a strong electric field can be generated inside the plasma plume, and these fields can accelerate the ions up to MeV energies in the case of \ced{a laser with} high power density. The possibility to install a small accelerator based on non-equilibrium expanding plasmas is under investigation in many laboratories, and also at LNS some interesting results have been obtained with low-power-density lasers, accelerating ions up to 10{\u}keV \cite{torrisi}.

If the plasmas produced by laser ablation are used as ion sources, the performance summarized in Table \ref{LIS} can be \ced{achieved}.

\begin{table}[!htbp]
\caption{Characteristic performance of laser ion sources.}\label{LIS}
\centering
\begin{tabular}{cccc}
 \hline\hline
 \textbf{Charge state} &\textbf{Power density} &\textbf{Current} &\textbf{Pulse time}\\
 & \textbf{(W{\u}cm$^{-2}$)} & \textbf{(mA)} & \textbf{($\rmmu$s)} \\
 \hline
 \noalign{\vspace{3pt}}
 1--2        & $10^9$        & 10--100 & 20 \\[3pt]
 $z\geq10^+$ & $\geq10^{12}$ & 0.5--10 & 5--10\\[3pt]
 \hline\hline
\end{tabular}
\end{table}

\section{Electron cyclotron resonance ion sources (ECRIS)}

The ECR ion sources are characterized by high-level technology; and the use of superconducting magnets and high-frequency microwave generators has became mandatory for third-generation sources. The general layout of these sources can be summarized as in Fig.~\ref{fig,ECRIS_subsystems}.

\begin{figure}[htbp]
\begin{center}%\pspicture (-1,0)(6,7.5)
\includegraphics[width=0.98\textwidth]{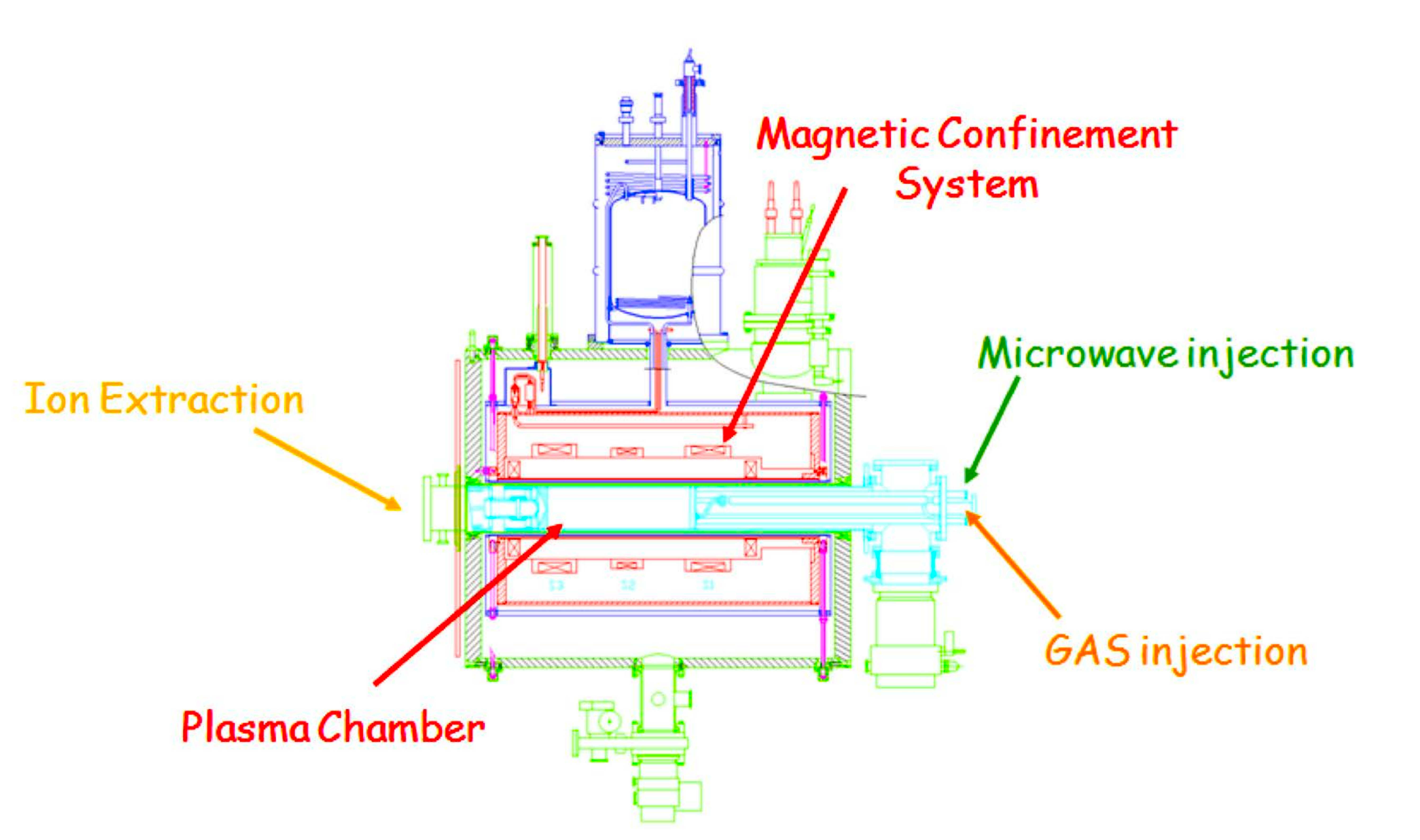}
\end{center}%\endpspicture
\caption{\label{fig,ECRIS_subsystems} Main subsystems of the ECRIS}
\end{figure}

ECR ion sources are based on the so-called electron cyclotron resonance, which ensures resonant absorption of electromagnetic energy by the plasma electrons. \ced{To this end}, the plasma must be created in the presence of a magnetostatic field appositely shaped to guarantee effective plasma confinement.
In brief, an ECRIS works with a gas or vapour injected into a vacuum chamber, with microwaves feeding the cavity and in the presence of an appositely shaped magnetic field. A few free electrons spiralling around the magnetic field lines are accelerated (and then heated) by the microwaves according to the process called electron cyclotron resonance \cite{gell}. The electron--atom collisions allow multiply charged ions to be obtained, and very high charge states can be reached because of the quite long confinement times.

Here we can summarize the advantages and disadvantages of such machines:
\begin{itemize}
     \item  \textit{Advantages}\\
     Intense currents even for highly charged ions\\
     Long source lifetime\\
     High stability\\
     Capability to produce CW and pulsed beams
     \item  \textit{Disadvantages}\\
     High power consumption\\
     Expensive and complex technology both for magnets and for microwave components\\ High ripple for highly charged ion beams (up to $\sim 10\%$)\\
     Long conditioning times\\
     Some difficulties for the production of metallic beams
\end{itemize}

\subsection{ECRIS magnets technology}

The different subsystems that compose the ECR ion sources are shown in Fig.~\ref{fig,ECRIS_subsystems}.
An efficient confinement system is mandatory to obtain high-density and stable plasmas. It consists of two or more solenoids for the axial confinement and a hexapole for the radial confinement. The dynamics of the plasma electrons along the plasma chamber axis can be described on the basis of simple mirror system characteristics. In order to obtain the minimum $B$ (min-B) configuration, which ensures magnetohydrodynamic (MHD) stability, a sextupole encloses the plasma chamber, creating a field increasing monotonically along the chamber radius.
Figure \ref{fig,magneticstructure} illustrates the design of an ECRIS.

\begin{figure}[htbp]
\begin{center}
\includegraphics[width=0.75\textwidth]{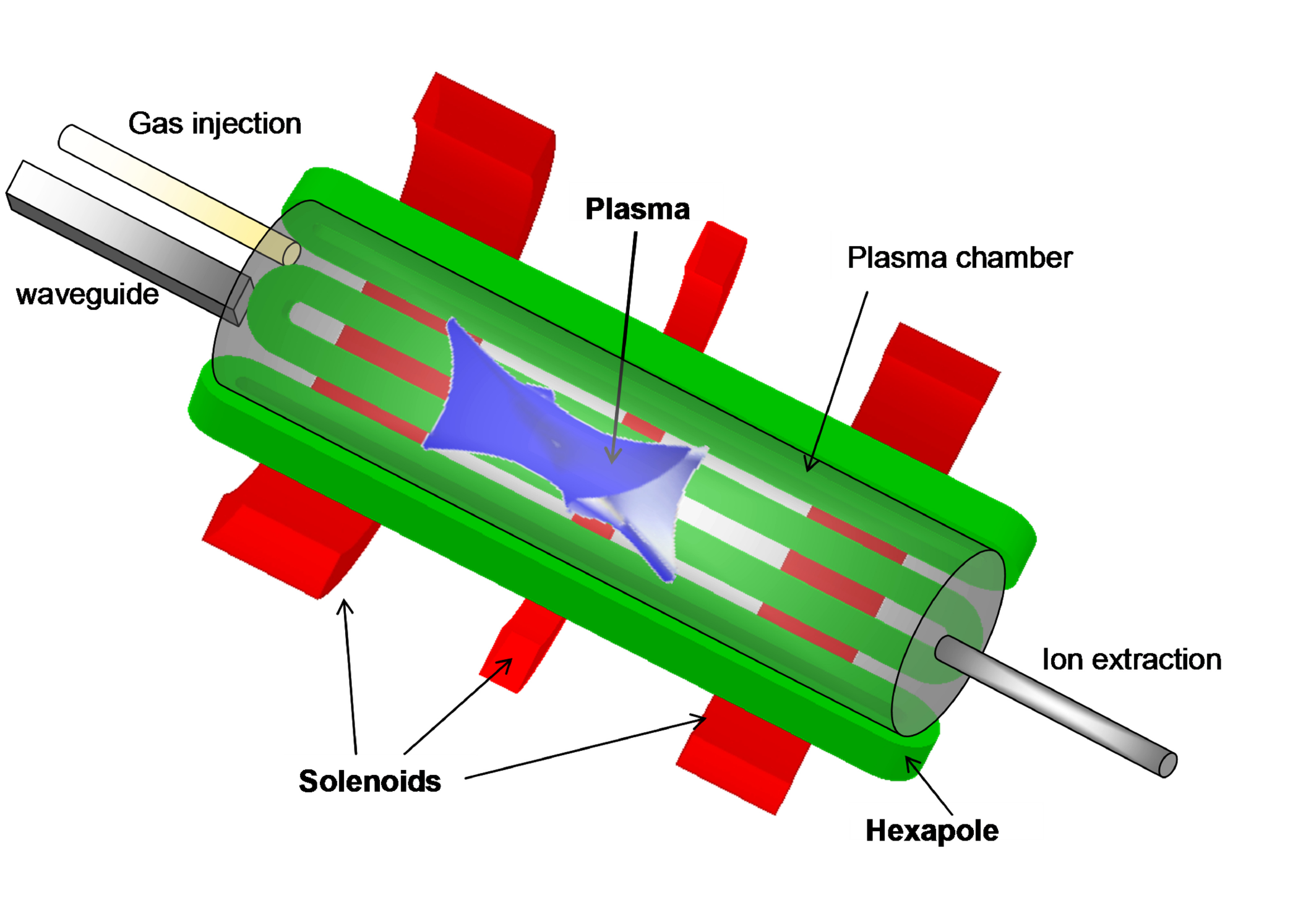}
\end{center}
\caption{\label{fig,magneticstructure} Location of the main ECRIS devices (magnets, microwave injection, gas injection, ion extraction, plasma chamber) together with the six-cusp plasma generated in the central region of the plasma chamber.}
\end{figure}

The technological problems related to the magnetic systems used for ECRIS are not simple to solve, because of the very high values of the magnetic field needed for the source operations. In order to give an estimation of the magnetic field needed for operations at high frequencies, we report the expression giving the strength of the magnetic field at a fixed microwave frequency:
\begin{equation}
B_{\rm ECR}=2\pi\frac{m}{q}f=0.03577f \quad (\mathrm{GHz}).
\end{equation}
Hence, if $f=18${\u}GHz, then $B_{\rm ECR}\simeq0.64${\u}T.
Obviously this value of the magnetic field \ced{concerns} the resonance, but to ensure high stability and long confinement times the magnetic field must be considerably higher. In third-generation ECRIS (like the VENUS source operating at LNBL, California), magnetic fields up to 3.7{\u}T are used. Furthermore, the possibility to operate with variable magnetic field is highly appreciated, as this allows the source performance to be optimized. In general, several types of multipoles may be used, and it is known that the higher the multipole order, the higher will be the loss area of the magnetic structure.
Larger areas mean higher ion currents, but on the other hand it lowers the ion lifetime and thus the mean ion charge state (see Eq.~\eqref{eq,qmed_propto_Q}).
In order to reach a compromise between the extraction current and the ion lifetime, the hexapole is employed in the most of ECRIS. The loss area of the hexapole is shown in Fig.~\ref{fig,areaperditaesa}.

\begin{figure}[h]
\begin{center}
\includegraphics[width=0.90\textwidth]{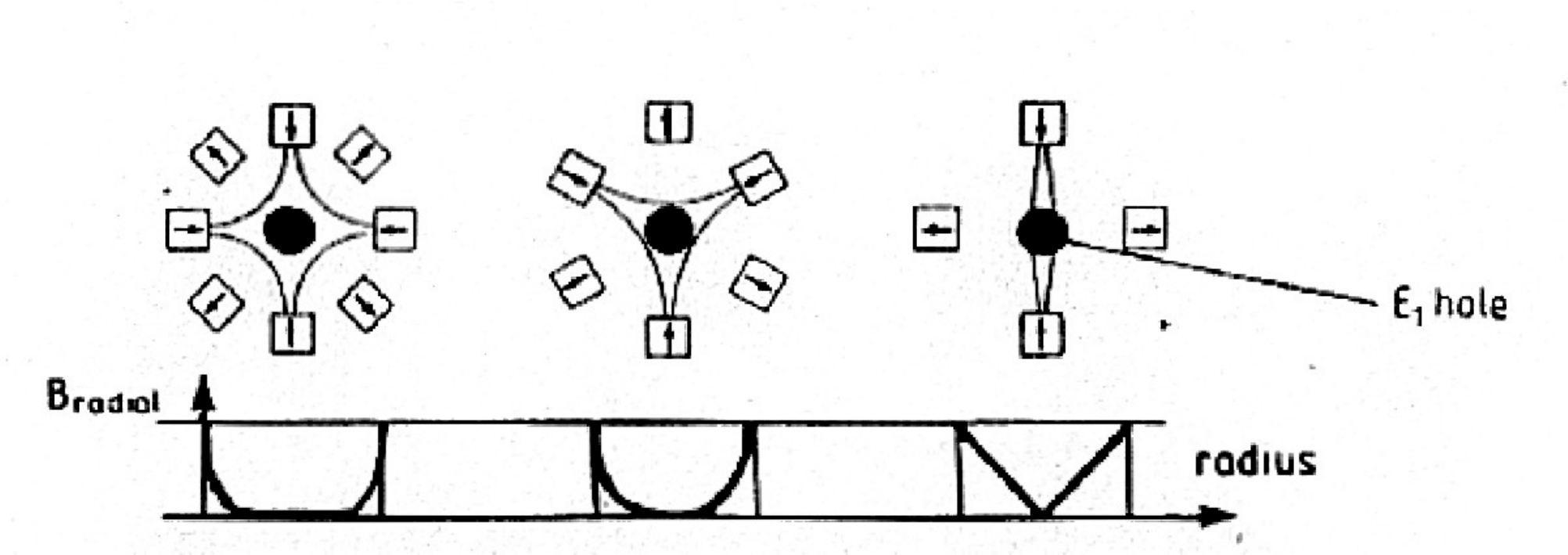}
\end{center}
\caption{\label{fig,areaperditaesa} Loss area and trend of the radial magnetic field in the case of multipolar magnets}
\end{figure}

Figure \ref{fig,areaperditaesa} shows the loss area shape for three different magnetic multipoles (octupole, hexapole and quadrupole). It also shows the radial trend of the magnetic field for each magnet: the field grows on going from the centre to the periphery. Let us define by $B_{\rm m}$ and $r_{\rm m}$ the magnetic field (on the multipole walls) and the radius of the multipole, and by $r_{\rm C}$ the plasma chamber radius. Then the magnetic field on the plasma chamber walls is given by
\begin{equation}
\label{eq,111}
B_{\rm C}=B_{\rm m}\left(\frac{r_{\rm C}}{r_{\rm m}}\right)^{\frac{1}{2}n-1},
\end{equation}
where $n$ corresponds to the multipole order. Note that the higher the multipole order, the lower will be the value of the magnetic field at the chamber walls, which is detrimental for the confinement. Hence the hexapole is the best compromise between the maximum value of the magnetic field and the loss area.
Figure \ref{fig,hexapolepicture} shows the superconducting hexapole of the MS-ECRIS source. In this case the maximum hexapolar field on the chamber wall should be 2.7{\u}T (the maximum available for ECRIS currently operating in the world).

\begin{figure}[htbp]
\begin{center}
\includegraphics[width=0.95\textwidth]{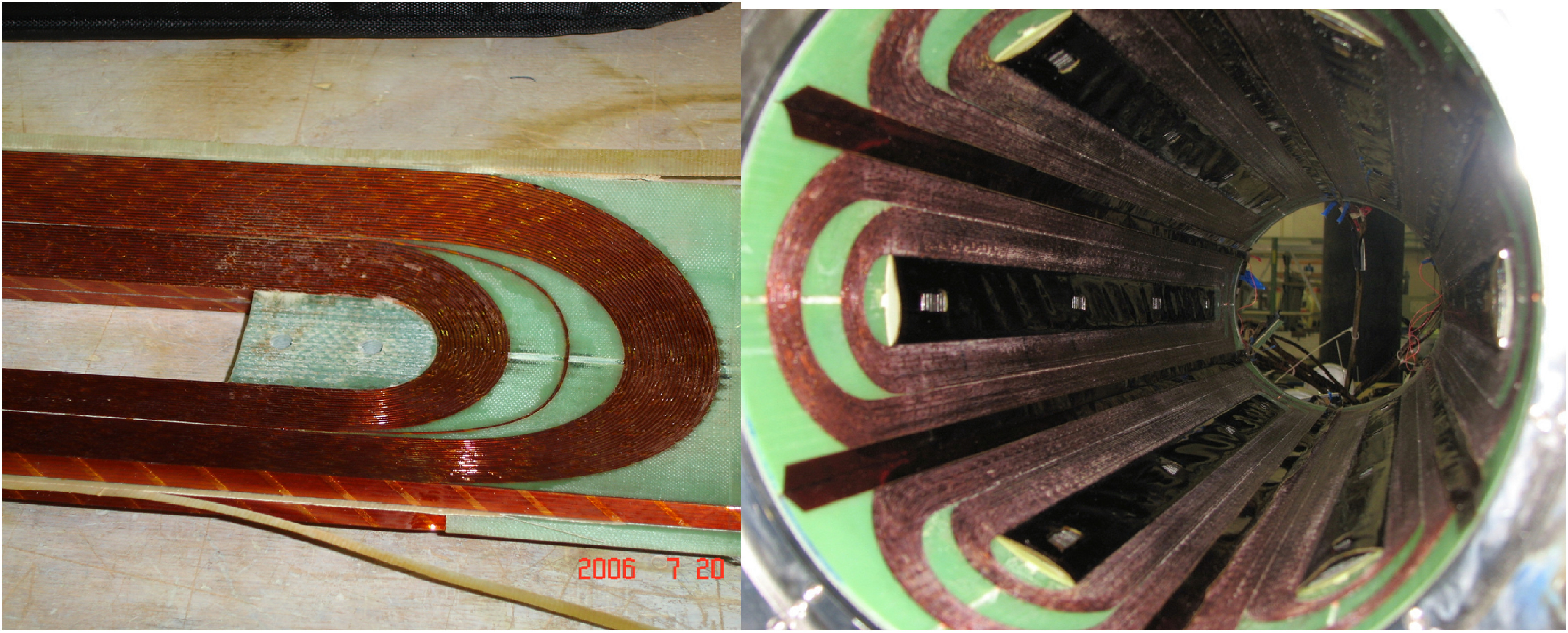}
\end{center}
\caption{\label{fig,hexapolepicture} Superconducting coils and hexapole assembly for the MS-ECRIS source}
\end{figure}

\subsection{The microwave injection system}

The microwave injection system consists of one or more waveguides with the corresponding ports located in the injection flange of the plasma chamber. As an example, Fig.~\ref{fig,microwaveinjectionSERSE} shows the injection flange with the location of the microwave ports for the SERSE source.

\begin{figure}[htpb]
\begin{center}
\includegraphics[width=0.85\textwidth]{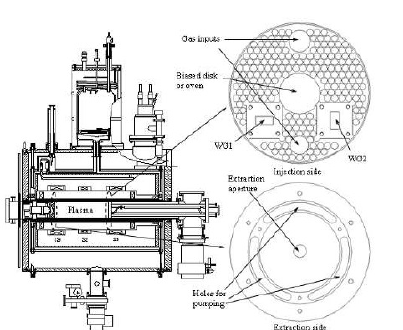}
\end{center}
\caption{\label{fig,microwaveinjectionSERSE} Detail of the injection flange of the SERSE source with the two microwave ports and a sketch of the microwave generators--plasma chamber connections via waveguides. }
\end{figure}

The insulation of the chamber is ensured by a DC break placed in proximity to the injection flange. When two waveguide ports are present, the two injection ports are often perpendicular to each other. The \textit{microwave generators} are usually of two types:  \textit{klystron generators} and  \textit{travelling wave tube (TWT) amplifiers}. For third-generation sources the gyrotron-based generators are used, with frequencies of 28{\u}GHz or higher.

\subsection{The plasma chamber}

The plasma chambers used for ECRIS usually consist of a cylindrical cavity with stainless steel or aluminium walls. As an example, the SERSE chamber is made of aluminium \ced{and has} a radius of 6.5{\u}cm and a length of 45{\u}cm. The volume of the MS-ECRIS source, designed for the GSI Laboratory, Darmstadt, is slightly bigger, with a chamber radius of 7.5{\u}cm and length of 50{\u}cm. The chamber injection flange includes the  \textit{gas injection system} together with the holes for vacuum pumping and the microwave ports. On the opposite side of the plasma chamber there is the  \textit{extraction system}. The chamber has a hole, which permits the extraction of the ions. It is also used for the  chamber vacuum pumping; it is usually biased up to high voltages (e.g.\ 25{\u}kV for SERSE). The extraction of intense beams is ensured by a two- or three-electrode system. Details about the extraction system of ECRIS can be found elsewhere \cite{gell}.

\subsection{Additional devices}

Additional devices are used for the ECRIS operations. A  \textit{solenoid} or other focusing elements are adopted after the extraction system.  \textit{Faraday cups} are used to measure the beam current; profilers and emittance measurement devices permit the beam quality to be defined. A magnetic dipole is used as charge-over-mass particle selector, and it permits the charge state distribution (CSD) to be determined. A pressure of the order of $1 \times 10^{-7}${\u}mbar can be reached by means of rotative and turbomolecular pumps.

The ECRIS are also able to obtain metallic ion beams, but a gaseous medium needs to be ignited and transformed into plasma through the electron cyclotron resonance. The production of metallic beams requires the presence of an  \textit{oven} that  can be installed into the plasma chamber. The oven can operate with different principles: resistive or inductively heated; and it is often able to reach a temperature higher than 2000$^\circ$C. Figure \ref{fig,oven} shows the inductively heated oven built for MS-ECRIS.

\begin{figure}[htbp]
\begin{center}
\includegraphics[width=0.7\textwidth]{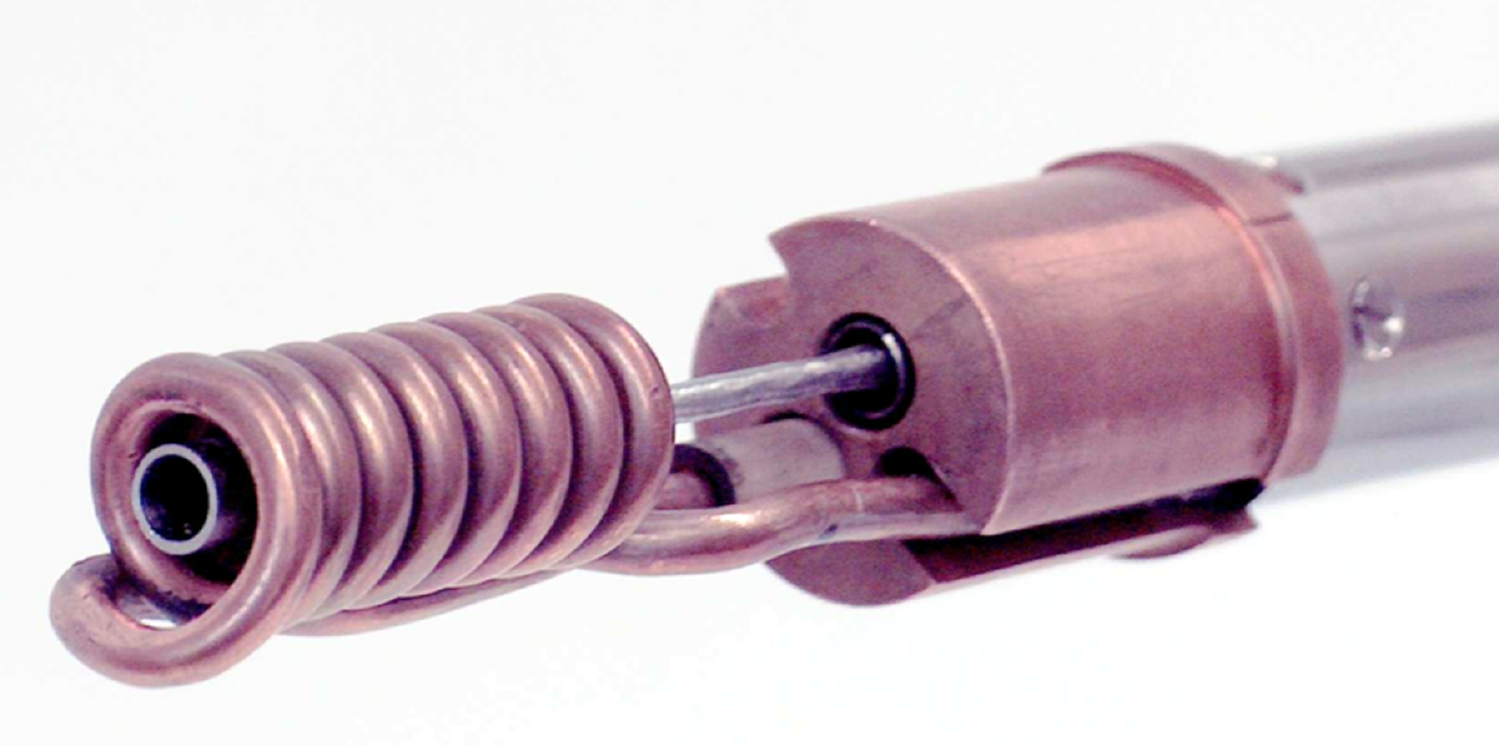}
\end{center}
\caption{\label{fig,oven} The inductively heated oven constructed for the MS-ECRIS source}
\end{figure}

Some problems arise when metals from refractory elements have to be used (like tungsten, tantalum, etc.). For these cases a new method named ECLISSE was developed at LNS; it couples the advantages of the ECRIS in terms of HCI beams and high current, low energy spread and low emittance, with the possibility to obtain ions of several metals by means of laser ablation. Such a method may represent a solution for HCI production from metals, as it may give very high currents of extremely high charge states, but it requires that the ECRIS plasma has a high energy content $n_{\rme}kT_\rme$.

In order to increase the electron density, additional tools have been studied, designed and tested over the past years. In particular the \textit{biased disc} is \ced{widely} used and it consists of a metallic and cylindrically shaped electrode inserted into the plasma and biased at different voltages (from a few tens to hundreds of volts). Also its position inside the cavity can be varied, but usually it stays outside the plasma core, that is, out of the region of the plasma inside the resonance surface. This device permits \ced{replacement of} the electrons lost because of confinement leakages. \ced{Thus a biased} electrode immersed into the plasma is able to supply additional electrons and to reduce the so-called `electron starvation' that limits the performances of ECRIS, by reducing the number of ionizing collisions inside the plasma core.

Another ancillary component for modern ECRIS with superconducting magnets is the cryostat containing liquid helium. Figure \ref{fig,MSECRIScryostat} shows the design of the MS-ECRIS cryostat  surrounding the plasma chamber. Owing to the complexity of ECRIS systems and to the parameters \ced{that need to be controlled} (as well as for safety \ced{reasons}), a remote control panel is usually installed, which \ced{in particular} allows the microwave generators' power (and frequency where possible) \ced{to be varied} and the magnetic field profile (in the case of variable magnetic field) \ced{to be changed}.

\begin{figure}[htbp]
\begin{center}
\includegraphics[width=0.7\textwidth]{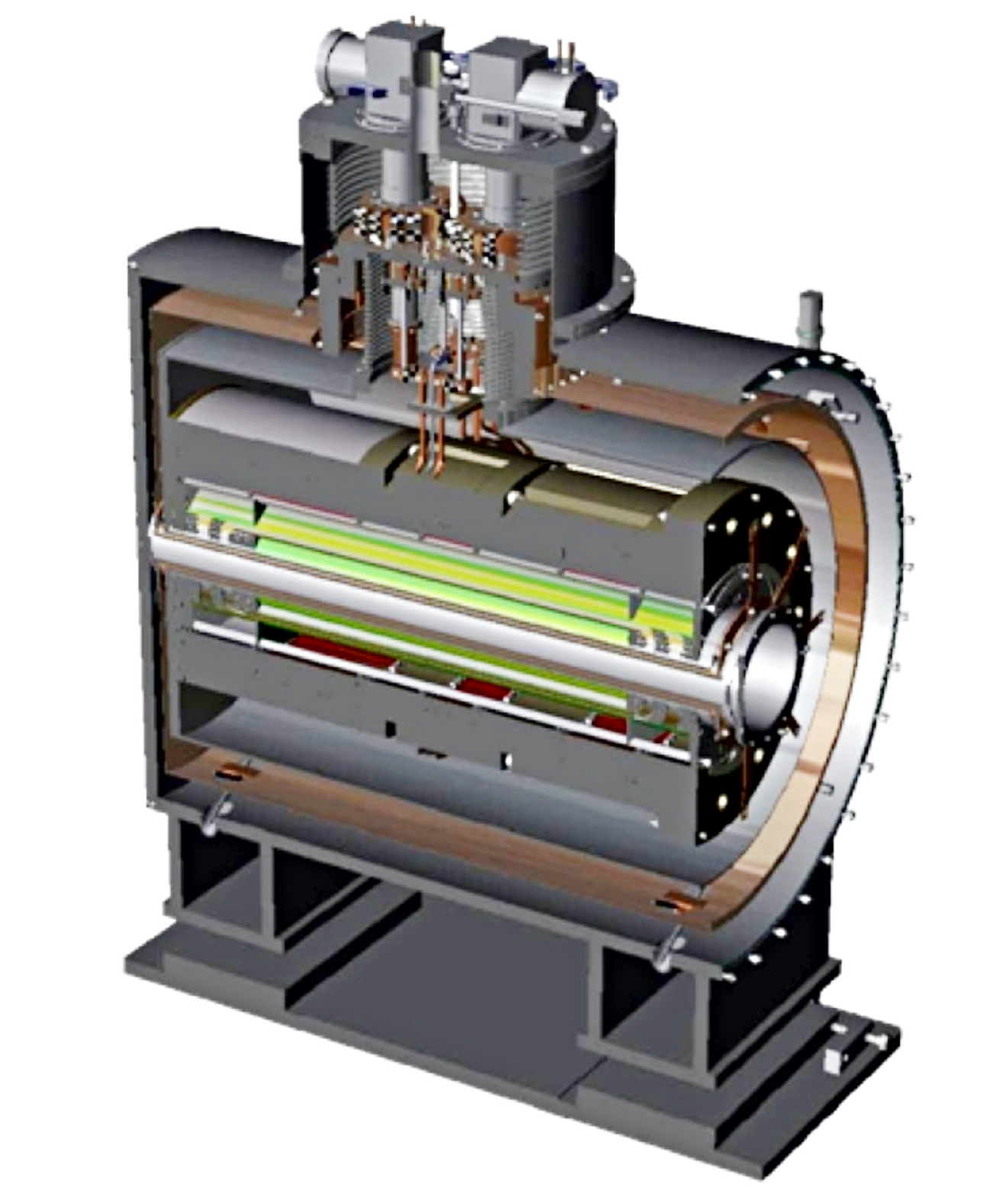}
\end{center}
\caption{\label{fig,MSECRIScryostat} The design of the MS-ECRIS cryostat and of the plasma chamber}
\end{figure}

\section{Fundamentals about magnetic confinement of plasmas}
\label{ref,magnetic_confinement}

Confinement of charged particles, and thus of plasmas, is of primary importance in many fields of physics. In the case of ion sources, the ions must live for milliseconds inside the plasma, so that they are subjected to a number of collisions.
Confinement can be ensured by electric or magnetic fields. While \ced{the effect of an electric field on} charged particles is quite trivial to discuss, the role of a non-uniform magnetic field is not. Magnetic confinement can be studied in either a single-particle or \ced{one-fluid} approximation.

Let us start by recalling the main equations of fluid dynamics as applied to plasmas.
The general equation that describes the momentum transport in a plasma subjected to electric and magnetic fields is
\begin{equation}
\label{eq,fluidmotionsenzacoll}
mn\left[\frac{\partial \vec{u}}{\partial t} + (\vec{u}\cdot\vec{\nabla})\vec{u}\right]
 = qn (\vec{E}+\vec{u}\times\vec{B})-\vec{\nabla}p .
\end{equation}
In the one-fluid approximation, i.e.\ when the fluid equations are written for a mixture of ions and electrons and the average mass density and velocity are calculated, we can make some approximations \ced{that considerably simplify} the model. We suppose that the particles move with velocities less than $c$ (as in the case of the single-particle approximation), we take into account the plasma quasi-neutrality (i.e.\ $n_\rme\simeq n_\rmi$), we neglect the ratio $m_\rme/m_\rmi$ and minor terms, we consider only long-spatial-scale phenomena, with the further restriction that only low-frequency phenomena occur, and finally we consider that the system remains isotropic at all times. Hence the set of so-called  \textit{magnetohydrodynamic} equations becomes
\begin{eqnarray}
\label{eq,MHD1}
\frac{\partial \rho_m}{\partial t}+\nabla\cdot\rho_m\vec{V}&=&0,\\
\label{eq,MHD2}
\rho_m\frac{\partial \vec{V}}{\partial t}&=&\frac{\vec{J}\times\vec{B}}{c}\nabla p,\\
\label{eq,MHD3}
\vec{E}+\frac{\vec{V}\times\vec{B}}{c}&=&\eta \vec{J},\\
\label{eq,MHD4}
\vec{\nabla}\times\vec{E}&=&-\frac{1}{c}\frac{\partial \vec{B}}{\partial t},\\
\label{eq,MHD5}
\vec{\nabla}\times\vec{B}&=&\frac{4\pi \vec{J}}{c}.
\end{eqnarray}
where the subscript "m" in the density term indicates ions or electrons respectively, since the equations are valid for both species.

\ced{From Eq.~\eqref{eq,MHD2}, for plasmas in the stationary state, we obtain}
\begin{equation}
\label{eq,condizioneEquilibrioMHD}
\vec{\nabla}p=\vec{J}\times\vec{B}.
\end{equation}
From Eq.~\eqref{eq,condizioneEquilibrioMHD} it follows that the currents and the magnetic field lines lie on isobaric surfaces, and the vectors $\vec{J}$ and $\vec{B}$ are orthogonal each other.

Furthermore, by making use of the equation
\begin{equation}
\label{eq,MaxwellMHD}
\vec{\nabla}\times\vec{B}=\mu_0\vec{j},
\end{equation}
Eq.~\eqref{eq,MHD1} becomes
\begin{equation}
\label{eq,definingmagneticpressure}
\vec{\nabla}\left(p+\frac{B^2}{2\mu_0}\right)
=\frac{1}{\mu_0}(\vec{B}\cdot\vec{\nabla})\vec{B}.
\end{equation}
Assuming that $\vec{B}$ varies weakly along its direction, we can write
\begin{equation}
\label{eq,sommapressMHD}
p+\frac{B^2}{2\mu_0}=\mathrm{constant}.
\end{equation}
In Eq.~\eqref{eq,sommapressMHD}
${B^2}/({2\mu_0})$ is the magnetic pressure. The sum of the kinetic pressure and the magnetic pressure is constant, and thus, in a plasma with a density gradient, the magnetic pressure will be high where the particle pressure is low (i.e.\ in the plasma periphery), and it will be low where the plasma density is high.
This diamagnetic effect is due to the so-called  \textit{diamagnetic current}, which arises as a consequence of the pressure gradient.
The intensity of the diamagnetic effect can be evaluated by \ced{taking the ratio of} the two terms in Eq.~\eqref{eq,sommapressMHD}. Such a ratio is usually indicated as the  \textit{$\beta$ parameter}:
\begin{equation}
\label{eq,betaMHD}
\beta\equiv\frac{\sum nkT}{{B^2}/{2\mu_0}}.
\end{equation}
Generally, $\beta$ can be used to determine the extent of the plasma confinement:
\begin{enumerate}
\item [(a)] for $\beta \ll 1$,  the magnetic pressure is much higher than the plasma kinetic pressure, which is the condition for an optimal confinement;
\item [(b)] for $\beta > 1$,  the plasma pressure is higher than that due to the magnetic field, and thus the plasma cannot be magnetically confined; and
\item [(c)] for $\beta = 1$,  the diamagnetic effect generates an internal magnetic field exactly equal to the external one, and thus two regions exist, one where  only the magnetic field is present, and the other where only the plasma, without magnetic field, exists.
\end{enumerate}
It can be demonstrated that only in case of $\beta \ll 1$ is the plasma well confined; otherwise the equilibrium is unstable.

The \ced{simplest} device for plasma confinement was investigated \ced{long ago} by Fermi and it is named a \textit{simple mirror}.\footnote{Fermi proposed the simple mirror configuration, with sources of the magnetic fields moving towards the centre of the system, as a possible mechanism involved in the generation of cosmic rays.} In this magnetic configuration, the field is provided by two solenoids with coinciding axes, located \ced{at definite distances from} other. Figure \ref{fig,mirror} features the shape of the field lines.

The principal constraints for charged particle confinement are summarized here:
\begin{enumerate}
    \item  a magnetic field gradient is needed $\vec{\nabla}_{\parallel}B$, directed along the direction of the field itself;
    \item  even in simple cases, as for simple mirror configurations, the field may be axisymmetric but it must have a radial component; and
    \item  the velocity vector of the charged particle must have a particular angle with the field lines; a boundary exists for this angle, and for angles smaller than this, the particle is no longer confined.
    \end{enumerate}

In Fig.~\ref{fig,mirror} only one of the two mirror elements is displayed, and the Larmor motions and the drift around the axis are shown.
Along the mirror axis $z$, charged particles experience a force $F_z$, which is able to confine them provided that some specific conditions are satisfied.

\begin{figure}[htbp]
\begin{center}
\includegraphics[width=0.6\textwidth]{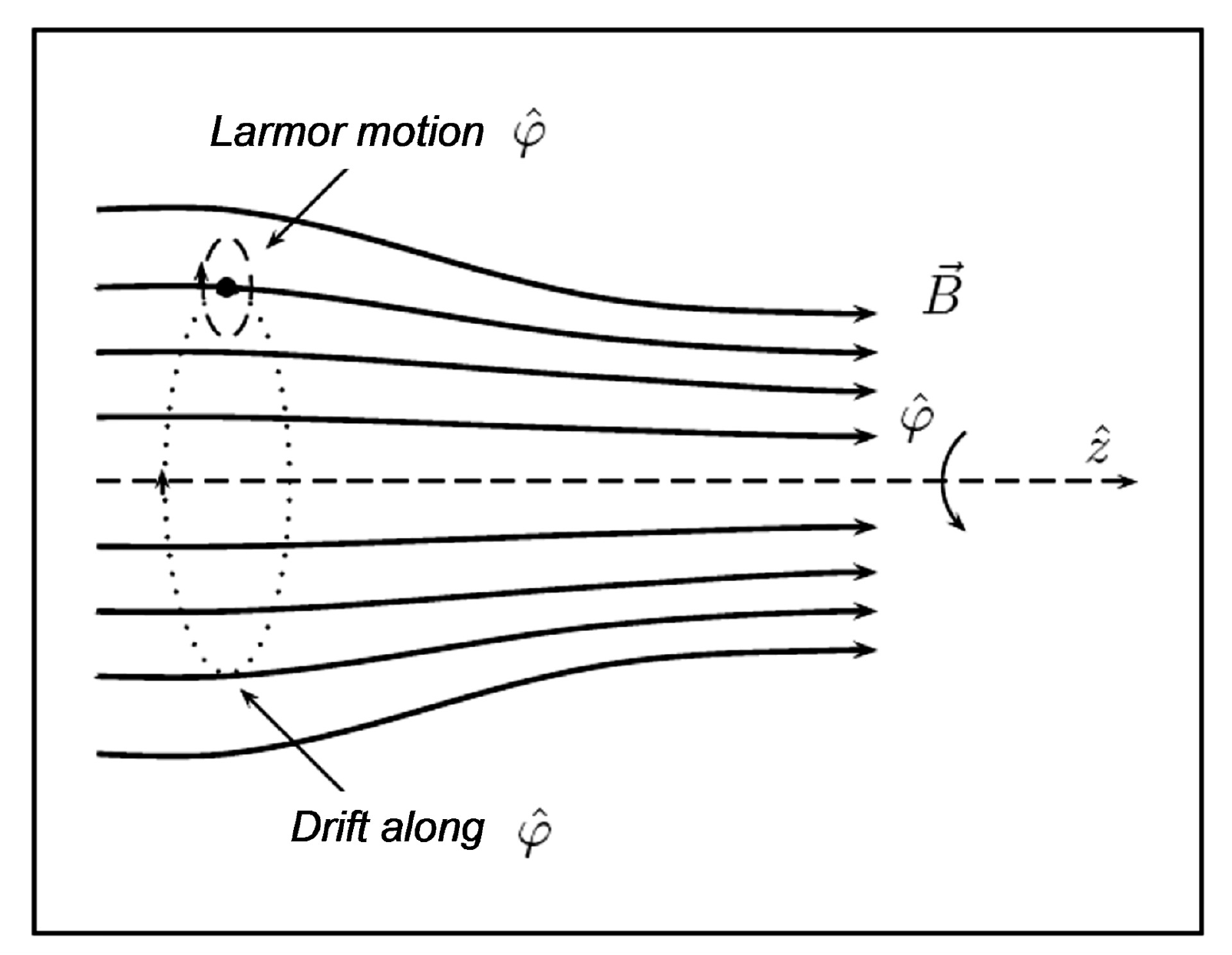}
\end{center}
\caption{\label{fig,mirror} Magnetic field lines produced in a simple mirror configuration. The density of the field lines increases from left to right, and this generates the gradient, which works as a mirror for charged particles.}
\end{figure}

The correct expression for $F_z$ is to be averaged over a gyro-period. For the sake of simplicity we consider a particle moving along the mirror axis. Then $v_{\varphi}$ is constant and depending on the sign of $q$ it will be $v_{\varphi}=\mp v_{\perp}$. Furthermore $r=r_{\rm L}$ and thus
\begin{equation}
\label{eq,Fzmedia}
\bar{F}_z=\mp\frac{1}{2}qv_{\perp}r_{\rm L}\frac{\partial
B}{\partial
z}=\mp\frac{1}{2}q\frac{v_{\perp}^2}{\omega_{\rm c}}\frac{\partial
B_z}{\partial z}=-\frac{1}{2}\frac{mv_{\perp}^2}{B}\frac{\partial
B_z}{\partial z}.
\end{equation}
By averaging $F_z$ over a gyro-period, considering a particle moving along
the mirror axis, we obtain \cite{chen}
\begin{equation}
\bar{F}_z=-\frac{mv_{\perp}^2}{B}B_1z,
\end{equation}
and thus the force is directed in the opposite direction with respect to the particle motion. We can also define the so-called  \textit{magnetic moment} of the charged particle as
\begin{equation}
\label{eq,momentomagnetico}
\mu\equiv\frac{1}{2}\frac{mv_{\perp}^2}{B}.
\end{equation}
Then Eq.~\eqref{eq,Fzmedia} becomes
\begin{equation}
\label{eq,Fzmu}
\bar{F}_z=-\mu\left(\frac{\partial B}{\partial z}\right).
\end{equation}
\ced{Generalizing this for whatever gradient occurs along the particle motion, we obtain}
\begin{equation}
\label{eq,Fzgenerale}
\vec{F}_{\parallel}=-\mu\frac{\partial B}{\partial
s}=-\mu\vec{\nabla}_{\parallel}B,
\end{equation}
where $\rmd\vec{s}$ is a line element along the $\vec{B}$ direction.

The main property of $\mu$ is that it is an adiabatic invariant for the particle motion.\footnote{To read more about the adiabatic invariants in plasmas, see Ref.~\cite{chen}}.
The confinement of charged particles in mirror-like devices can be studied in terms of $\mu$ invariance.
The breakdown of the adiabatic invariance of $\mu$ expels the particle from the confinement. The adiabatic invariance is valid as long as the $B_z$ component slowly varies with $z$ during a gyro-period. This condition can be written in the following manner:
\begin{equation}
\label{eq,condmodeling}
\frac{r_{\rm L}}{L}\ll1,
\end{equation}
i.e.\ the Larmor radius $r_{\rm L}$ must be much smaller than $L$, the characteristic length for the $B$ variation.\footnote{This property of the $\mu$ invariance is of primary importance also for ECRIS physics. If such a condition is not satisfied, then non-adiabatic effects begin to play a role, leading to the expulsion of the particle from the confinement.}

The mirror effect in terms of the $\mu$ invariance can be explained as follows.
If we consider a proof particle moving from the centre to the periphery of the magnetic bottle,\footnote{The simple mirror systems are often named `magnetic bottles'.} it sees a growing magnetic field. As $\mu$ must be kept constant, the perpendicular component of the velocity $v_{\perp}$ must increase because of the $B$ increase. The magnetic field does not \ced{do} any work on the system, and thus the kinetic energy remains constant. Hence when $v_{\perp}$ increases, $v_{\parallel}$ decreases. If $B$ is sufficiently high at the device periphery, there will be a mirror point where $v_{\parallel}=0$ and the particle is reflected.

\section{Ion confinement in ECR plasmas}
\label{sub,PotenzialTrap}

Magnetic confinement is effective only for particles that experience few collisions between the mirror points. This condition is often fulfilled for electrons but only rarely for ions. In ion sources, in fact, ions are cold, and their collisionality is a factor ${m_\rmi}/{m_\rme}$ larger than for electrons. This makes them poorly magnetized, especially in the plasma core. The conservation of plasma quasi-neutrality thereby makes the ion confinement a direct consequence of the electron one.
Presently, several models have been proposed to explain the ion confinement in magnetically confined plasmas, but only few experimental results are available to evaluate the best model.
The most commonly accepted model assumes that well-confined electrons in the plasma core provide a sufficiently strong negative potential dip that incorporates multicharged ions. This theory applies only to the plasma core, while for the whole volume a more general approach must be taken into account. The quasi-neutrality must be preserved everywhere, and an ambipolar diffusion process arises. The slower and the faster species can be determined by analysing the collisions times when collisions are responsible for particle deconfinement. The slower species determine the sign of the plasma potential.
Tonks and Langmuir formerly described the properties of plasmas located inside absorbing boundaries.\footnote{By absorbing boundaries we usually mean the walls of a metallic plasma chamber.} The preservation of the quasi-neutrality means that $n_\rme\simeq Zn_\rmi$, with $Z$ the mean ion charge state, and $j_\rme\simeq j_\rmi$. Thus the electron and the ion densities must remain the same, along with the currents of particles leaving the plasma. Starting from this assumption in ECRIS plasma halo, the electrons,\footnote{Because of their higher mobility, due to the low plasma halo temperature and its high density, the electrons are more mobile than ions in the plasma periphery, and so they tend to leave the plasma at a higher rate than ions.} having ${\tau_{\ei}}/{\tau_{\ii}}<1$,  leave the plasma more easily than ions, thus creating a positive potential at the plasma boundary, which accelerates ions and tends to retain the escaping electrons. On the contrary, in the plasma core, the electrons are less collisional, because their temperature is very high. Then they are well confined and generate a hot electron cloud, which retains the ions. The negative potential dip can be indicated as $\Delta\phi$:
\begin{equation}
\Delta\phi\propto-\frac{T_\rmi}{ze}.
\end{equation}
Assuming that the bouncing time of ions inside the plasma trap is
$\tau_{\rm b}$ we obtain
\begin{eqnarray}
\label{eq,bouncepotwell}
\tau_{\rm b} &\simeq& \frac{B_{\max}}{B_{\min}}\left[\frac{\mathrm{plasma\
length}}{\langle \mathrm{ion\ velocity} \rangle}\right], \\
\tau_{i} &\simeq& \exp\left(\frac{Z\Delta\phi}{KT_\rmi}\right),
\end{eqnarray}
with $Z=ze$. Note that the plasma potential is globally positive, but the `hollow shape', with the small dip in the plasma core, confines the ions.

According to this model, the electrons with high energies ($T_\rme>10${\u}keV) provide a stabilizing effect for the plasma because they practically confine the ions. In order to increase the ion lifetime, we have to increase the \ced{depth of the potential}.

This model for ion confinement has been recently put in \ced{doubt} by some results and needs further developments.

\subsection{Onset of MHD instabilities and stabilized min-B traps}

The simple mirror configuration allows \ced{confinement of those particles that obey the} adiabatic invariance of the magnetic moment. Gibson, Jordan and Lauer proved experimentally that singly charged particles remain trapped in a magnetic mirror in a manner completely consistent with the prediction of the adiabatic theory. Thus this theory predicted (and the experiment confirmed) that single particles are satisfactorily trapped in a magnetic mirror field. However, the trapping of a single particle in some field configuration does not mean that a plasma has a stable equilibrium. Not only losses induced by collisions (which can change the velocity vector direction, putting it inside the loss cone) but also other instabilities may arise that increase particle losses. Because of the magnetic confinement, the simple mirror plasmas have a non-isotropic energy distribution function. This leads to some instabilities in the velocity space, the so-called  \textit{loss cone instability}. However, some hydrodynamic instabilities may be more dangerous. The most important instability involving simple mirror plasmas is the so-called  \textit{flute instability}. This instability for magnetically confined plasmas is equivalent to the well-known Rayleigh--Taylor instability, which arises in neutral fluids and also in plasmas subjected to the gravitational force. In the case of magnetically confined plasmas, the curvature of magnetic field lines triggers the instability. In Fig.~\ref{fig,minBconf} a schematic representation of the mechanism at the basis of the flute instability is reported.

\begin{figure}[htpb]
\begin{center}
\includegraphics[width=0.75\textwidth]{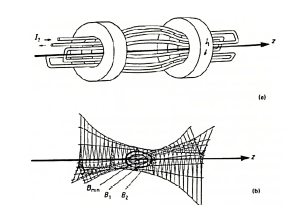}
\end{center}
\caption{\label{fig,minBconf} (a) Magnetic system and (b) magnetic field originated by two solenoids surrounding a hexapole}
\end{figure}

If the magnetic field lines curve into the plasma everywhere, then the confinement will be stable. This result gives us some practical information to realize optimal plasma traps where plasmas are stably magnetically confined.

In 1962 Ioffe \cite{io} first reported on plasma-confinement experiments in a magnetic field configuration that had the desirable feature that the magnetic field increased in every direction away from the plasma boundary, and that did not have the undesirable feature of a region where the magnetic field went to zero inside the plasma (as in the case of cusp configurations), producing an open wide loss cone.
Such a configuration can be obtained as a superposition of two magnetic field structures: one created by two solenoids (simple mirror), and the other created by six conductors surrounding the plasma chamber (a hexapole). In this way the magnetic field increases in every direction that can be seen from the plasma centre. In fact, the field produced by the hexapole increases along the plasma radius. This provides a stabilizing effect also radially, the simple mirror already being able to confine the plasma axially.
This configuration can also be thought of as a multi-mirror device: the particle is reflected at many points travelling along the field lines. Ioffe and his co-workers at the Kurchatov Institute in Moscow used a device known as the PR-5 (Probkotron-5) to demonstrate that the hydromagnetic instability could be suppressed with a min-B configuration. Figure \ref{fig,minBconf} schematically shows the field lines of such a field.

Note that the geometrical loci where  $B$ is constant are ellipsoids that are usually defined as \textit{egg-shaped surfaces}.
As there are no axial symmetries, the $\mu$ adiabatic invariance cannot be invoked to study particle confinement.
 Another adiabatic invariant can be used for min-B configurations, i.e.
\begin{equation}
\label{eq,secondoinvadiab}
J=\oint v_{\parallel}\, \rmd l,
\end{equation}
in which $J$ essentially determines the length of field lines between two reflection points. The $J$ invariance implies that, even after reflection, the particle will move along the same field line, or at least along a field line with the same $J$ value.
Field lines with the same $J$ value define a surface over which particles with fixed values of ${W}/{\mu}$ will move.

The system is realized so that the constant-$J$ surfaces do not intercept the plasma chamber walls. Up to now we have pointed out the ability of a min-B configuration in ensuring the MHD stability, but no numbers have been given for the relative values of the confining magnetic field, i.e.\ for the mirror ratio at the extraction and injection sides, and also in the radial direction.

Gammino and Ciavola proposed, in 1990, the so-called  \textit{high-B mode} (HBM) concept \cite{RSI}, stating that  `only an appropriate optimization, in terms of mirror ratios, of the confining magnetic field, leads to the exploitation of the electron density as expected from the frequency scaling, thus leading to an effective increase of the sources performances'. Such a principle does not conflict with the Geller laws, but it limits their applications to well-confined plasmas.
The main aspect of the HBM concerns the possibility to give some numbers for the configuration of the magnetic field. These results have been obtained by virtue of MHD considerations. Note that a stable MHD equilibrium can be reached only if the magnetic pressure is higher than the plasma pressure.\footnote{By plasma pressure we mean the usual kinetic pressure due to the particle temperature and density.} Then we estimate the plasma pressure in ECRIS, we compare it with the magnetic pressure (which depends on the maximum field that can be produced by the magnet), and finally we can state whether or not a given magnetic configuration is stable according to the MHD criteria for stability.\footnote{Magnetically confined plasmas can be considered MHD-stable only if the $\beta$ parameter is appreciably smaller than one.}

The expression for the $\beta$ parameter is
\begin{equation}
\label{eq,betaconf}
\beta=\frac{n_\rme KT_\rme}{{B^2}/{2\mu_0}}.
\end{equation}
Well-confined plasmas are characterized by \ced{$\beta$ parameters in the range}
\begin{equation}
\label{eq,condbeta}
0.005<\beta<0.01.
\end{equation}
The plasma electron density can be roughly estimated as
\begin{equation}
\label{eq,cutoffMHD}
n_\rme \simeq n_{\rm cutoff}
=\frac{m_\rme}{e^2}\epsilon_0\omega_{\rm RF}^2=\frac{\epsilon_0}{m_\rme}B^2_{\rm ECR},
\end{equation}
with
\begin{equation}
\omega_{\rm RF}=\frac{qB_{\rm ECR}}{m}.
\end{equation}
Hence, by substituting Eq.~\eqref{eq,cutoffMHD} into
Eq.~\eqref{eq,betaconf}, and considering Eq.~\eqref{eq,condbeta}, \ced{we find}
\begin{eqnarray}
\frac{\epsilon_0}{m_\rme}B^2_{\rm ECR}kT_\rme &<& 0.01\frac{B^2}{2\mu_0},\\
\left(\frac{B}{B_{\rm ECR}}\right)^2 &>& 100 k T_{\rme}2\frac{\mu_0\epsilon_0}{m_\rme}.
\end{eqnarray}
We also have
\begin{equation}
\mu_0\epsilon_0=\frac{1}{c^2} \quad \mbox{and} \quad m_\rme c^2=511{\u}\mathrm{keV},
\end{equation}
\ced{so that we then obtain}
\begin{equation}
\label{eq,HBM}
\left(\frac{B}{B_{\rm ECR}}\right)^2 > \frac{200kT_\rme}{511{\u}\mathrm{keV}}.
\end{equation}
The above equation is of primary importance for the operations of modern ECRIS, as it gives some useful information on practical preparation of an experiment.
For instance, if $kT_\rme=10${\u}keV, then
\begin{equation}
\label{eq,condizionecampo}
\frac{B}{B_{\rm ECR}}>2.
\end{equation}
These results have been confirmed by a great amount of experimental data. One \ced{example} is shown in Fig.~\ref{fig,HBM_exp_confirmation}, showing the trend of the extracted current from the SERSE source of the INFN-LNS for different values of the radial mirror ratio ${B}/{B_{\rm ECR}}$. Note that the extracted current rapidly increases, then tends to saturate when ${B}/{B_{\rm ECR}}>2$, confirming that the electron density reaches its maximum value only when the plasma stability condition is established.
\begin{figure}[htbp]
\begin{center}
\includegraphics[width=0.70\textwidth]{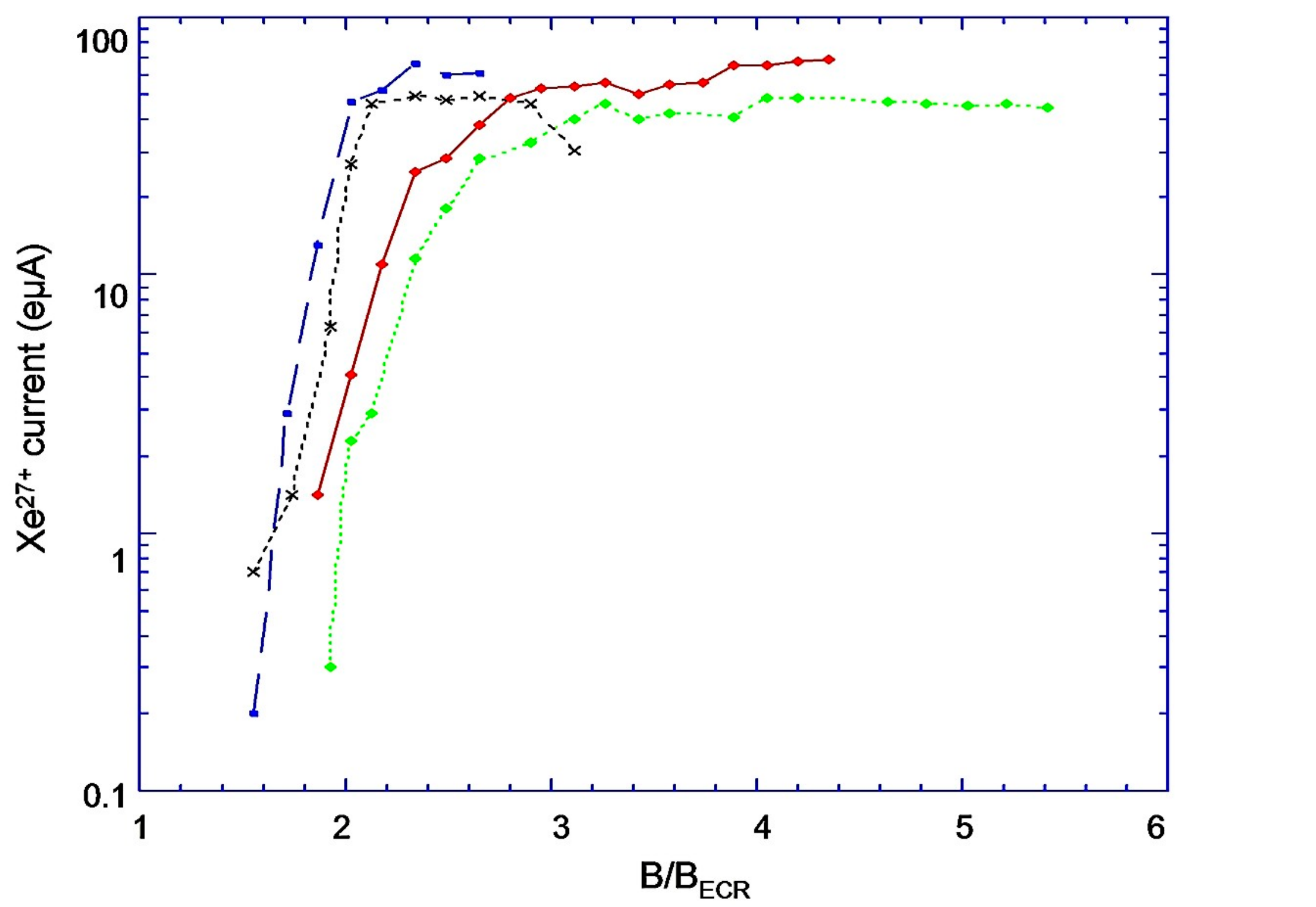}
\end{center}
\caption{\label{fig,HBM_exp_confirmation} Experimental confirmation of the HBM concept. The extracted current increases as the ratio $B/B_{\rm ECR}$ increases, but it saturates for $B/B_{\rm ECR}=2$, in good agreement with Eq.~\eqref{eq,HBM} in the case of $T_\rme\simeq10${\u}keV.}
\end{figure}
More precise scaling laws can be determined by taking into account all the typical values of the magnetic field in the plasma trap: the injection, the extraction and the radial field, all of them compared with the field at ECR and with that at the bottom of the configuration (usually named $B_{\min}$ or $B_0$).
Finally, note that the relation \eqref{eq,HBM} strictly links the magnetic configuration to the used frequency ($B_{\rm ECR}$ is fixed by the microwave frequency). Hence,  to effectively improve the ECRIS performance, one has to increase both the microwave frequency and the magnetic field, in particular the mirror ratios, of the plasma trap.

\section{Simplified treatment of wave propagation in a magnetized plasma}

In strongly magnetized plasmas the direction of the external magnetic field is a cause of anisotropy in the propagation of the electromagnetic waves. The electron motion is in fact not free in any direction, and waves are sustained in different manners according to their propagation direction. Skipping all the theoretical details, let us here introduce the propagation constant also in the case of a generic angle of propagation:
\begin{equation}
\label{7.2.57}
k'_{\theta} = {\omega \over c} \left[1-{X(1+\rmi Z-X) \over
(1+\rmi Z) (1+\rmi Z-X) - \tfrac{1}{2}Y^2_{\rm T} +
\sqrt{\tfrac{1}{4}Y^4_{\rm T} + Y^2_{\rm L}(1+\rmi Z-X)^2}}\right]^{{1/2}}
\end{equation}
and
\begin{equation}
\label{7.2.58}
k''_{\theta} = {\omega \over c} \left[1-{X(1+\rmi Z-X) \over
(1+\rmi Z) (1+\rmi Z-X) - \tfrac{1}{2}Y^2_{\rm T} -
\sqrt{\tfrac{1}{4}Y^4_{\rm T} + Y^2_{\rm L}(1+\rmi Z-X)^2}}\right]^{{1}/{2}}.
\end{equation}

Hence we have found two waves travelling along an arbitrary direction defined by
$\theta$ with $k'_\theta$ and $k''_\theta$. As a function of $X$, $k'_\theta$ is more similar to a wave propagating in an isotropic plasma, and thus the associated wave will be called the  \textit{ordinary wave}; the wave with $k''_\theta$ is named  the \textit{extraordinary wave}. However, in plasma physics another classification prevails, according to the scheme shown in \ced{Figs.~\ref{fig,waves_propagation_scheme} and \ref{fig,RLOX}}.
Usually, the waves propagating along the magnetic field direction are called the $R$ and $L$ \textit{waves}, according to their polarization.\footnote{As may be argued by the notation, the $R$ wave is a right-hand polarized wave, and $L$ is a left-hand polarized wave.} The waves propagating along the perpendicular direction to $\vec{B}_0$ are named $O$ and $X$, according to the orientation of the wave electric field with respect to the magnetostatic field direction.
If $\vec{k}\perp\vec{B}_0$ and $\vec{E}\parallel\vec{B}_0$, then we have the $O$ mode. Otherwise, if $\vec{k}\perp\vec{B}_0$ and $\vec{E}\perp\vec{B}_0$, whatever the direction of $\vec{E}$, the wave will be called the $X$ mode.

\begin{figure}[htbp]
\begin{center}
\includegraphics[width=0.53\textwidth]{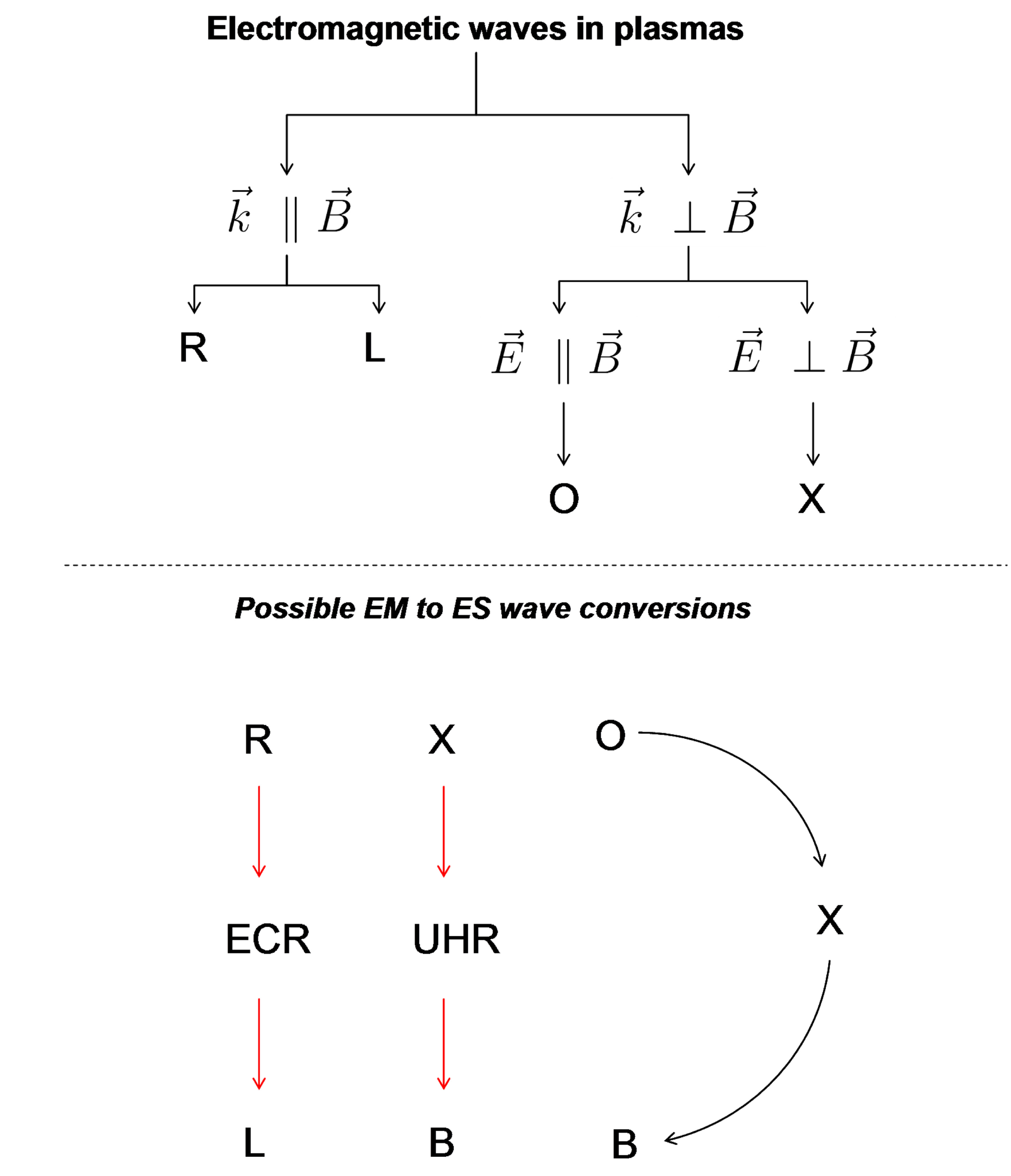}
\end{center}
\caption{ \label{fig,waves_propagation_scheme} Scheme of the wave propagation in anisotropic plasmas}
\end{figure}

\begin{figure}[htbp]
\begin{center}
\includegraphics[width=0.7\textwidth]{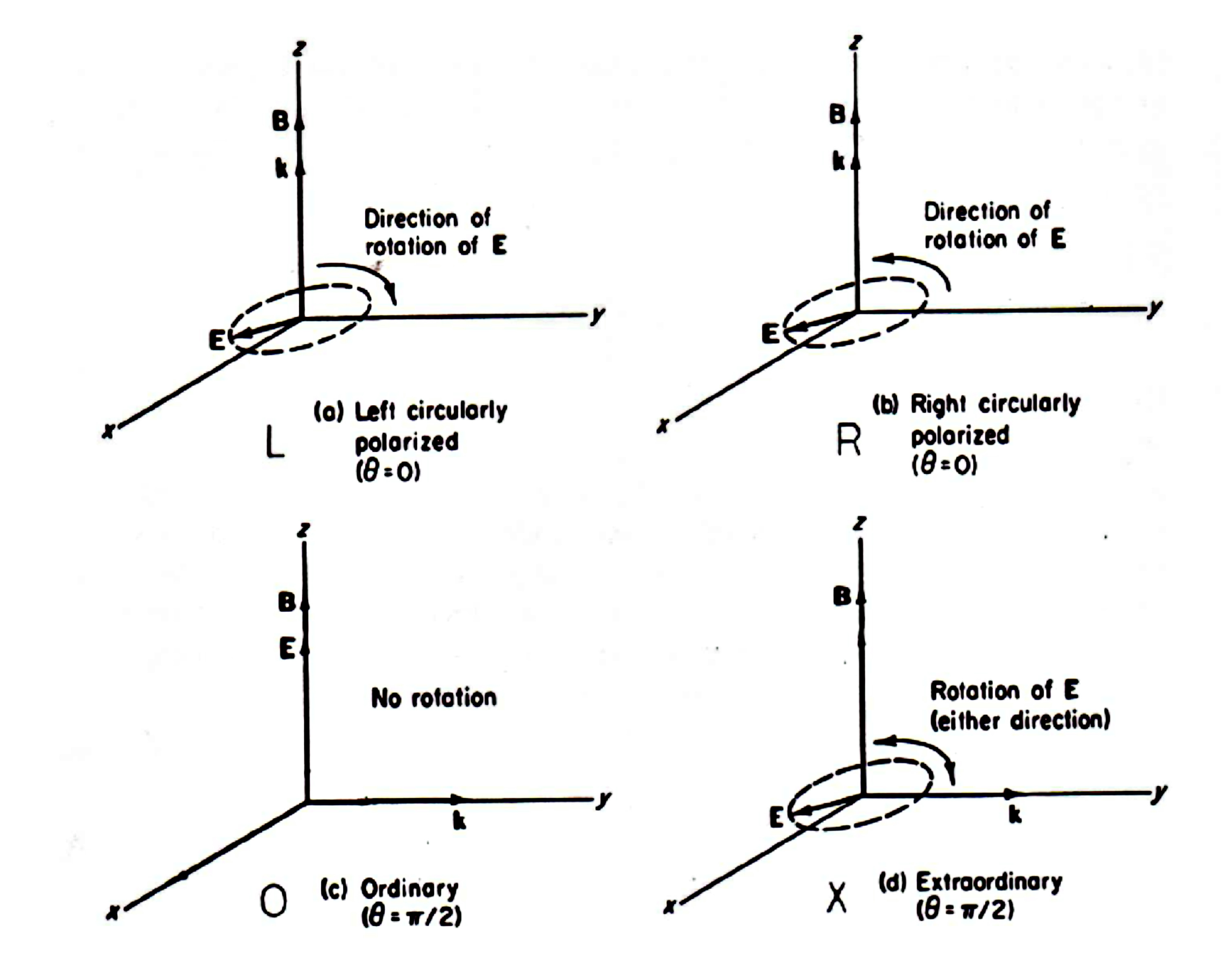}
\end{center}
\caption{\label{fig,RLOX} Diagram showing the possible orientations of the electric field with respect to the magnetostatic field, and also the possible polarizations in waves propagating in magnetized plasmas.}
\end{figure}

\begin{figure}[htbp]
\begin{center}
\includegraphics[width=0.7\textwidth]{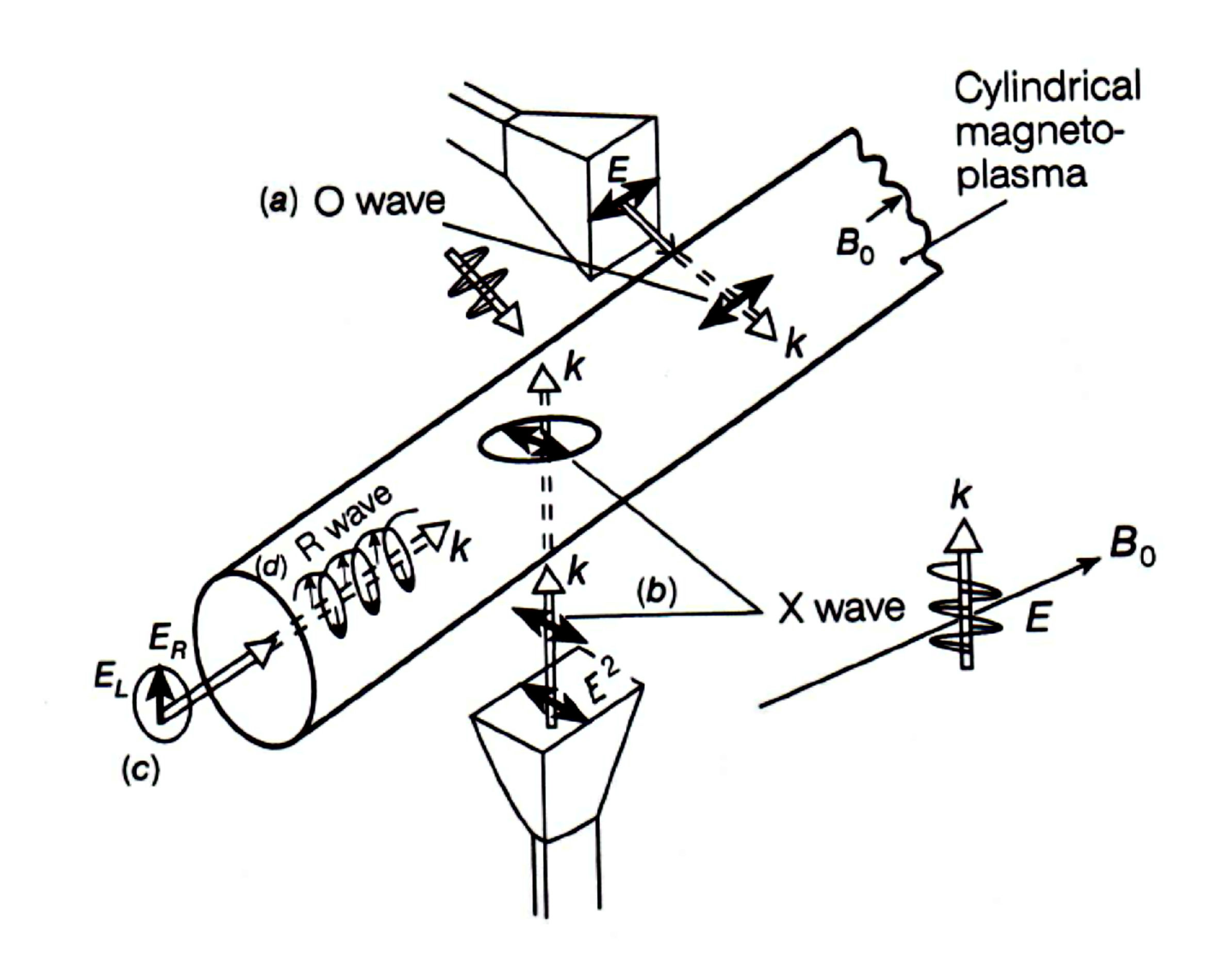}
\end{center}
\caption{\label{fig,RLOXinjection} Schematic representation of the possible ways to inject electromagnetic waves into a column of a magnetized plasma. The different modes can be excited, according to the location of the waveguide providing the electromagnetic field, and also according to the wave polarization inside the waveguide.}
\end{figure}

Many other waves can be generated in plasmas, many of them involving ions, collective fluid motions, and so on. \ced{The various ways in which electromagnetic waves can be injected into a column of magnetized plasma are shown in Fig.~\ref{fig,RLOXinjection}.}
By applying the kinetic theory, one can also determine the wave propagation in a warm plasma.

We next consider the cutoffs and resonances of the $R$, $L$, $O$ and $X$ waves. For the $R$ waves, $k$ becomes infinite at $\omega=\omega_{\rm c}$. In this case the microwave frequency is equal to the Larmor frequency, hence the wave is in resonance with the cyclotron motion of the electrons. This resonance is called \textit{electron cyclotron resonance} and is of primary importance for ECRIS. More generally, the cutoffs and the resonances of the various propagation modes can be summarized by reporting the different ${c^2k^2}/{\omega^2}$ formulas obtained from Eqs.~\eqref{7.2.57} and \eqref{7.2.58}. A schematic summary of the main electron waves\footnote{\ced{The term `electron waves' usually indicates the high-frequency waves}, i.e. the electromagnetic waves. For these waves, the ions can be considered as an immobile medium, as the wave frequency is too high to have some ions responding to the electromagnetic field. It is usually assumed that ions ensure the quasi-neutrality, but they do not take part in the electromagnetic wave propagation.} is given in Table~\ref{tab,cutoffs_and_resonances}.

\begin{table}[!htbp]
\caption{Cutoffs and resonances of the waves propagating inside the plasma \cite{chen}.
\label{tab,cutoffs_and_resonances}}
\begin{center}
\begin{tabular}{lcc}
\hline\hline
\textbf{Vector orientation}&\textbf{Refractive index}&\textbf{Wave type}\\
\hline
\noalign{\vspace{3pt}}
  $\vec{B}_0=0$ & $\omega^2=\omega_{\rm p}^2+k^2c^2$ & Light waves
  \\[6pt]
  $\vec{k} \perp \vec{B}_0,\ \vec{E} \parallel \vec{B}_0$ & $\displaystyle\frac{c^2k^2}{\omega^2}
        = 1-\displaystyle\frac{\omega_{\rm p}^2}{\omega^2}$ & $O$ wave
  \\[12pt]
  $\vec{k}\perp\vec{B}_0,\ \vec{E}\perp\vec{B}_0$ & $\displaystyle\frac{c^2k^2}{\omega^2}
  =  1-\displaystyle\frac{\omega_{\rm p}^2}{\omega^2}
  \displaystyle\frac{\omega^2-\omega_{\rm p}^2}{\omega^2-\omega_{\rm h}^2}$  & $X$ wave
  \\[12pt]
  $\vec{k}\parallel\vec{B}_0$ & $\displaystyle\frac{c^2k^2}{\omega^2}
  = 1-\displaystyle\frac{\omega_{\rm p}^2/\omega^2}{1-(\omega_{\rm c}/\omega)}$ & $R$ wave \\[12pt]
  $\vec{k}\parallel\vec{B}_0$ &  $\displaystyle\frac{c^2k^2}{\omega^2}
  = 1-\displaystyle\frac{\omega_{\rm p}^2/\omega^2}{1+(\omega_{\rm c}/\omega)}$ & $L$ wave \\[9pt]
\hline\hline
\end{tabular}
\end{center}
\end{table}

In Table~\ref{tab,cutoffs_and_resonances}, $\omega_{\rm h}=\sqrt{\omega_{\rm c}^2 +\omega_{\rm p}^2}$ is the so-called  \textit{upper hybrid frequency}. It corresponds practically to the frequency of the Langmuir oscillations in non-magnetized plasmas. Electrostatic electron waves across $\vec{B}$ have this frequency, while those along $\vec{B}$ have the usual plasma oscillations with $\omega=\omega_{\rm p}$.
Furthermore, from Table \ref{tab,cutoffs_and_resonances} it can be seen that the $R$ wave suffers the resonance reported before (i.e.\ the ECR resonance), as its index of refraction goes to infinity when $\omega\rightarrow\omega_{\rm c}$. Also the $X$ mode has a resonance, but at the upper hybrid frequency. Thus, inside a magnetized plasma, where an $X$ mode is travelling, the so-called  \textit{upper hybrid resonance} (UHR) occurs if
\begin{equation}
\label{eq,UHR}
\omega^2=\omega_{\rm p}^2+\omega_{\rm c}^2=\omega_{\rm h}^2.
\end{equation}

The $L$ wave does not suffer any resonance, as well as the $O$ mode. Instead, they suffer a cutoff that can be determined by the equations reported in Table \ref{tab,cutoffs_and_resonances} when the index of refraction, i.e.\  ${c^2k^2}/{\omega^2}$, goes to zero. A powerful method to visualize the cutoffs and the resonances of the various modes is to plot them on a so-called Clemmow--Mullaly--Allis (CMA) diagram.\footnote{The CMA diagram takes its name from the three authors who proposed the graphical representation of the many cutoffs and resonances of electromagnetic waves propagating in a cold plasma.} This diagram is reported in Fig.~\ref{fig,CMA_1}.

\begin{figure}[htbp]
\begin{center}
\includegraphics[width=0.7\textwidth]{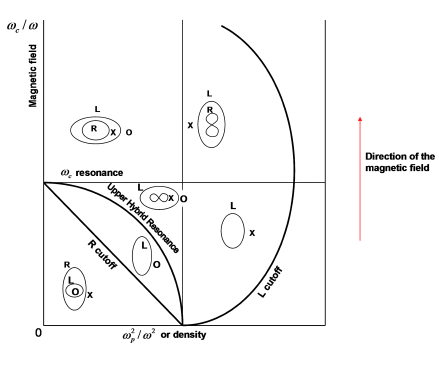}
\end{center}
\caption{\label{fig,CMA_1} The Clemmow--Mullaly--Allis diagram for the classification of waves in a cold plasma}
\end{figure}

When the propagation occurs at a given angle with respect to the magnetic field, the phase velocities change with the angle. Some of the modes listed above with  $\vec{k}\parallel\vec{B}_0$ or  $\vec{k}\perp\vec{B}_0$ change \ced{continually} into each other. Such a complicated picture is greatly simplified by the CMA diagram, which however works only in the case of the cold plasma approximation ($T_\rme=T_\rmi=0$). Any finite-temperature modification implies such a great number of complications that the diagram becomes completely \ced{useless}.

The CMA diagram can be viewed as a plot of $\omega_{\rm c}/\omega$ versus $\omega_{\rm p}^2/\omega^2$, or equivalently a plot of the magnetic field versus the plasma density. For a given frequency $\omega$, any experimental situation characterized by $\omega_{\rm c}$ (i.e.\ the magnetic field) and by $\omega_{\rm p}$ (i.e.\ the plasma density) is denoted by a point on the graph. The total plane is divided into several zones, and the boundaries of each zone are the cutoffs and the resonances mentioned above. For example, the upper hybrid resonance can be easily found in the graph. Considering an $X$ wave propagating from a region with high magnetic field, inside a plasma with a fixed value of density (e.g.\ $\omega_{\rm p}^2/\omega^2=0.7$), the value of the magnetic field where the UHR occurs can be easily determined on the graph.

\begin{figure}[htbp]
\begin{center}
\includegraphics[width=0.7\textwidth]{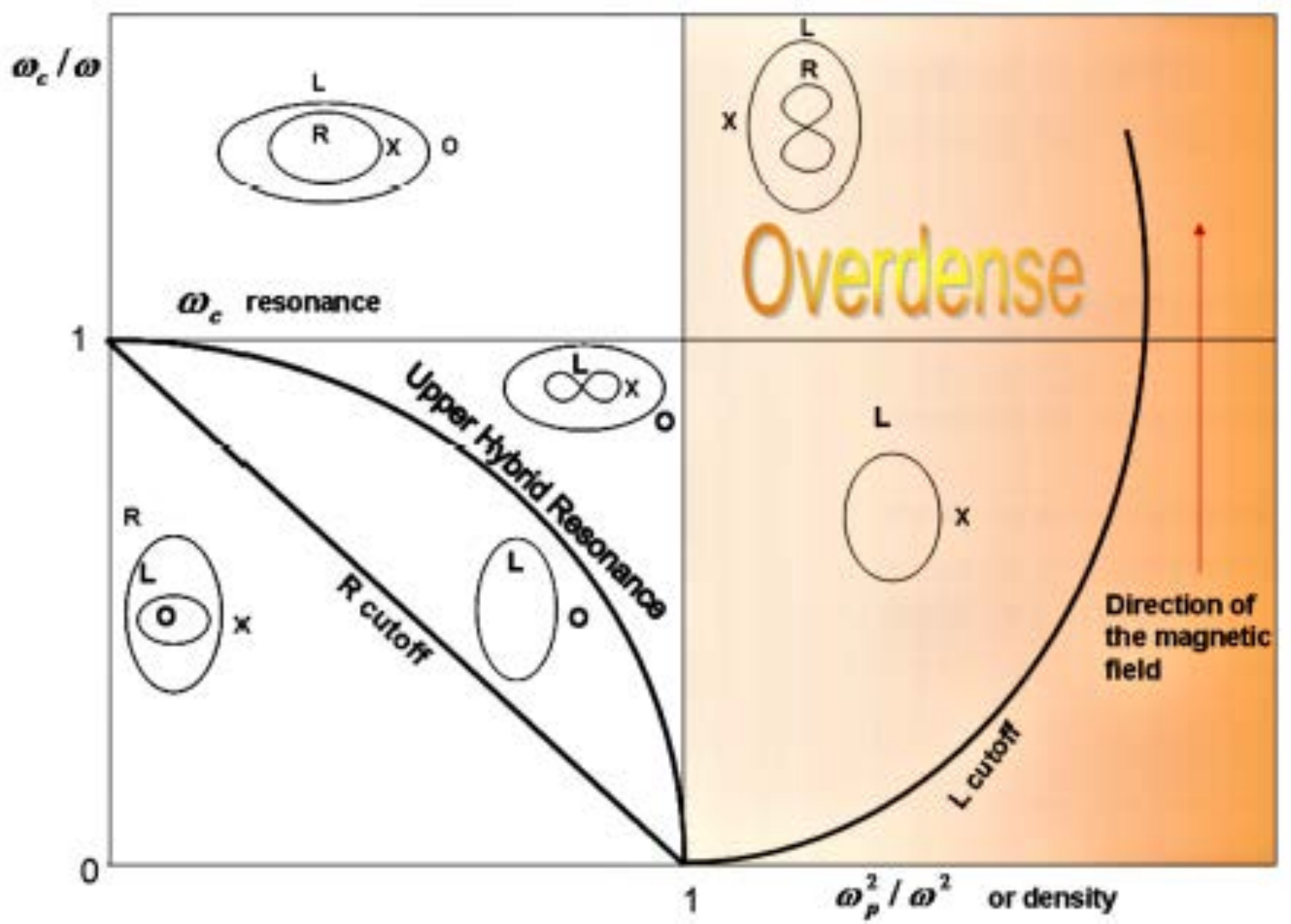}
\end{center}
\caption{\label{fig,CMA_overdense} A detailed view of the Clemmow--Mullaly--Allis diagram showing the region of the so-called overdense plasma, i.e.\ that region where the plasma density exceeds the cutoff of the ordinary ($O$) mode.}
\end{figure}

The vertical line at $\omega_{\rm p}^2/\omega^2=1$ is the so-called $O$ cutoff, and it corresponds to the cutoff density in an unmagnetized plasma. Above this value, the plasma will be called  \textit{overdense}, as shown in Fig.~\ref{fig,CMA_overdense}.
The small diagrams in each region delimited by the cutoffs and the resonances (i.e.\ the  ellipsoidal small diagrams) indicate not only which wave propagates, but also the qualitative variation of the phase velocity with the angle (considering that the magnetic field is directed along the vertical direction).

As mentioned above, the kinetic theory provides more information about the wave propagation than the macroscopic \ced{one-fluid} approach and the single-particle investigation. The main assumption of both \ced{the one-fluid} and single-particle approaches is that no velocity distributions of plasma particles are taken into account. This limits the applicability of such approaches to low-temperature plasmas only, or to those plasmas for which the particle velocities are near to the fluid element velocity.
Landau damping, ion waves and all those waves at harmonics of the cyclotron frequency are outside of the fluid or single-particle schemes, and they all depend on the plasma temperature. \ced{For example, in the case of the $O$ mode, whose dispersion relation is reported above and was based on a single-particle approach, if we apply the kinetic theory we find that other possible propagation modes exist in proximity of the cyclotron harmonics $n\omega_{\rm c}$.}

\section{Plasma heating by resonant stochastic wave--electron interaction}

In ECRIS an effective heating is experimentally observed; on the other hand, we know that the electron lifetimes are of the order of milliseconds for well-confined plasmas. Thus, non-adiabatic effects, which provide the stochasticity, arise in times shorter than milliseconds. Reasonable estimations and numerical simulations give a time of the order of a few nanoseconds to observe non-adiabatic effects in such plasmas. Thus, collisionless plasma heating is completely determined by the stochasticity, as demonstrated also by Liebermann and Lichtenberg \cite{LichtembergII}.

According to their theory, the stochasticity in ECR plasmas is provided by  multiple passages through the resonance, as long as a `forgetting' mechanism works well.
In order to better understand how the heating develops in multi-passage interactions, it is better to introduce the  \textit{separatrix} concept. Particles that resonantly interact with waves can be considered as trapped by the wave potential. The resonant condition is achieved if the well-known equation $\omega=kv_\rme$ is satisfied. Again, consider a frame moving with the wave velocity. Particles moving with a similar velocity can oscillate in the wave potential\footnote{As before, we consider a one-dimensional model to explain the particle trapping in the wave field at the resonance.} $\varphi(z)$. Then the electron moves with a velocity $v_\rme$, while the wave potential, which can be indicated by $\varphi(z)$,
\begin{equation}
\label{eq,potseparatrix}
\varphi=\varphi_0\sin(\omega t-kz),
\end{equation}
is stationary and sinusoidal, as the frame moves with $v_{\varphi}$. Because of energy conservation, we can write
\begin{equation}
\label{eq,conservazione_energia}
\tfrac{1}{2}mv_\rme^2+e\varphi(z)=\epsilon=\mathrm{constant}.
\end{equation}
Thus the velocity $v$ of the electron is
\begin{equation}
\label{eq,possible_vel_electron}
v_\rme=\pm\sqrt{\frac{2\epsilon-2e\varphi(z)}{m}}.
\end{equation}
This velocity can be plotted in the phase space $z$--$v_\rme$.

Particles at the bottom of the wave potential (i.e.\ perfectly resonant with the wave) stay exactly at the centre of each separatrix, and their trajectory in the phase space is just a point (their velocity with respect to the wave is zero, so they must lie on the $z$ axis). If an electron is moved from the bottom of the potential, it will begin to oscillate around the equilibrium point, similarly to mechanical oscillators with small-amplitude oscillations around the equilibrium. As long as the perturbation is small enough, the motion around the minimum will be harmonic. In the phase space, these motions are represented by ellipsoidal trajectories, as shown in Fig.~\ref{fig,separatrix_definition}. The electron velocity may become higher and higher, and then the ellipsoidal trajectories become bigger and bigger. In fact, if the electron energy becomes appreciable with respect to $\varphi(z)$, the motion is no longer harmonic and nonlinear effects arise. If the electron stays on the top of the $\varphi(z)$ potential, then the electron velocity will be zero according to Eq.~\eqref{eq,possible_vel_electron}. In this case the electron has the same velocity as the wave, but, because it stays on top of the wave potential, its equilibrium is not stable, and chaotic motions may arise. When this condition is satisfied, the phase space separates into different zones. Inside each zone the particle is trapped by the wave. Figure \ref{fig,separatrix_definition} features the separatrix of a sinusoidal wave interacting with a single particle. This theory, named the  \textit{pendulum model}, explains easily how the nonlinearity and the stochastic effects can play a fundamental role in the collisionless wave--particle interaction.
Those electrons whose velocity differs considerably from the wave phase velocity are not trapped and their velocity is slightly perturbed by the wave.
If the separatrix areas present some non-uniformities, for example, because of localized wave patterns, the heating becomes more effective, as such systems have other sources of stochasticity.

\begin{figure}[htbp]
\begin{center}
\includegraphics[width=0.7\textwidth]{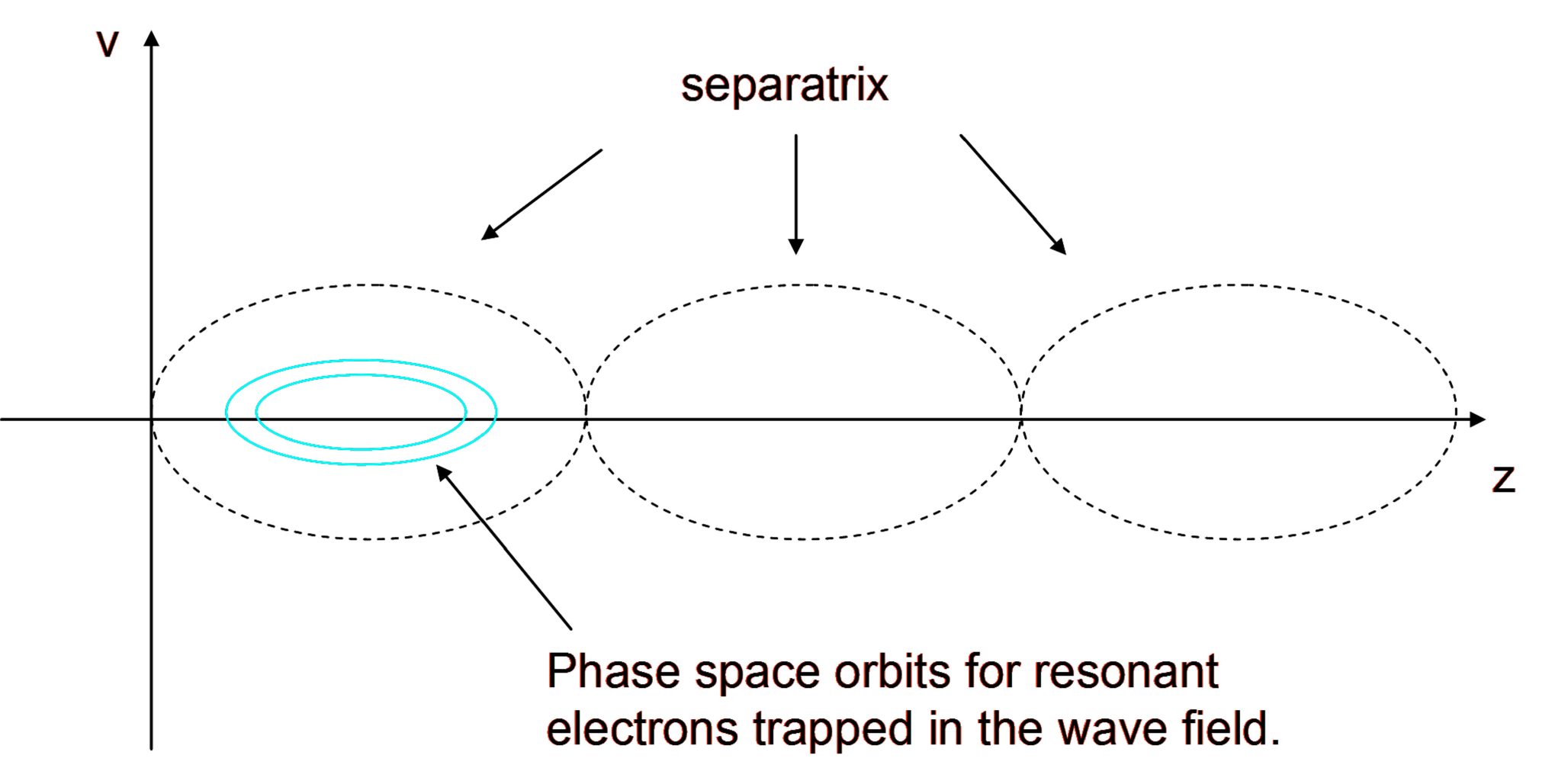}
\end{center}
\caption{\label{fig,separatrix_definition} Spread in the phase space of the electron velocity due to the activation of the stochastic effects}
\end{figure}

Let us discuss the heating in ECRIS through electron multi-passages across the ECR zones. An easier treatment is possible if we consider a simple mirror configuration, because in that case a one-dimensional model can be applied and the bouncing of the electrons inside the trap is subjected only to the potential $\mu B$. Then the equation for the $z$ variables becomes
\begin{equation}
\label{eq,zmotion}
z=z_0\cos(\omega_{\rm b}t+\psi_0),
\end{equation}
where $\omega_{\rm b}$ is the bouncing frequency. The perpendicular component of the wave electric field can be simply written as
\begin{equation}
\label{eq,perpelectricfield}
E_{\perp}(t,z,r,\theta)=E_{\perp{0}}\cos(\omega t-kz+\theta_{\perp}),
\end{equation}
where $\theta_{\perp}$ is the initial perpendicular phase. By substituting Eq.~\eqref{eq,zmotion} into Eq.~\eqref{eq,perpelectricfield}, we obtain
\begin{eqnarray}
\label{eq,composedmotion}
E_{\perp}(t,z,r,\theta)
&=&E_{\perp{0}}\cos[\omega t+\theta_{\perp}-kz_0\cos(\omega_{\rm b}t+\psi_0)]
    \nonumber \\
&=&\sum_n A \exp{[(\omega-\omega_{\rm c})t+\theta_{\perp}+\theta_{b}]},
\end{eqnarray}
where
\begin{equation}
\theta_{\rm b}=\omega_{\rm b}t+\psi_0.
\end{equation}

Equation \eqref{eq,composedmotion} takes into account the cyclotron frequency, the frequency of the wave and the bouncing frequency. The Fourier expansion \ced{shows} that the composition of the bounce and cyclotron frequencies leads to an effective multiwave interaction, i.e.\ we may imagine a single particle interacting with a \ced{pair} of waves at the resonance. Each wave--particle interaction has a proper separatrix and many zones like those shown in Fig.~\ref{fig,separatrix_definition} are present in the phase space.\footnote{Here we consider the perpendicular phase space, i.e.\ that one formed by $p_\perp$--$\theta_{\perp}$.} Figure \ref{fig,separatrix_overlap} shows the structure of several \ced{separatrices} in the phase space in the proximity of the ECR zone.

\begin{figure}[htbp]
\begin{center}
\includegraphics[width=0.75\textwidth]{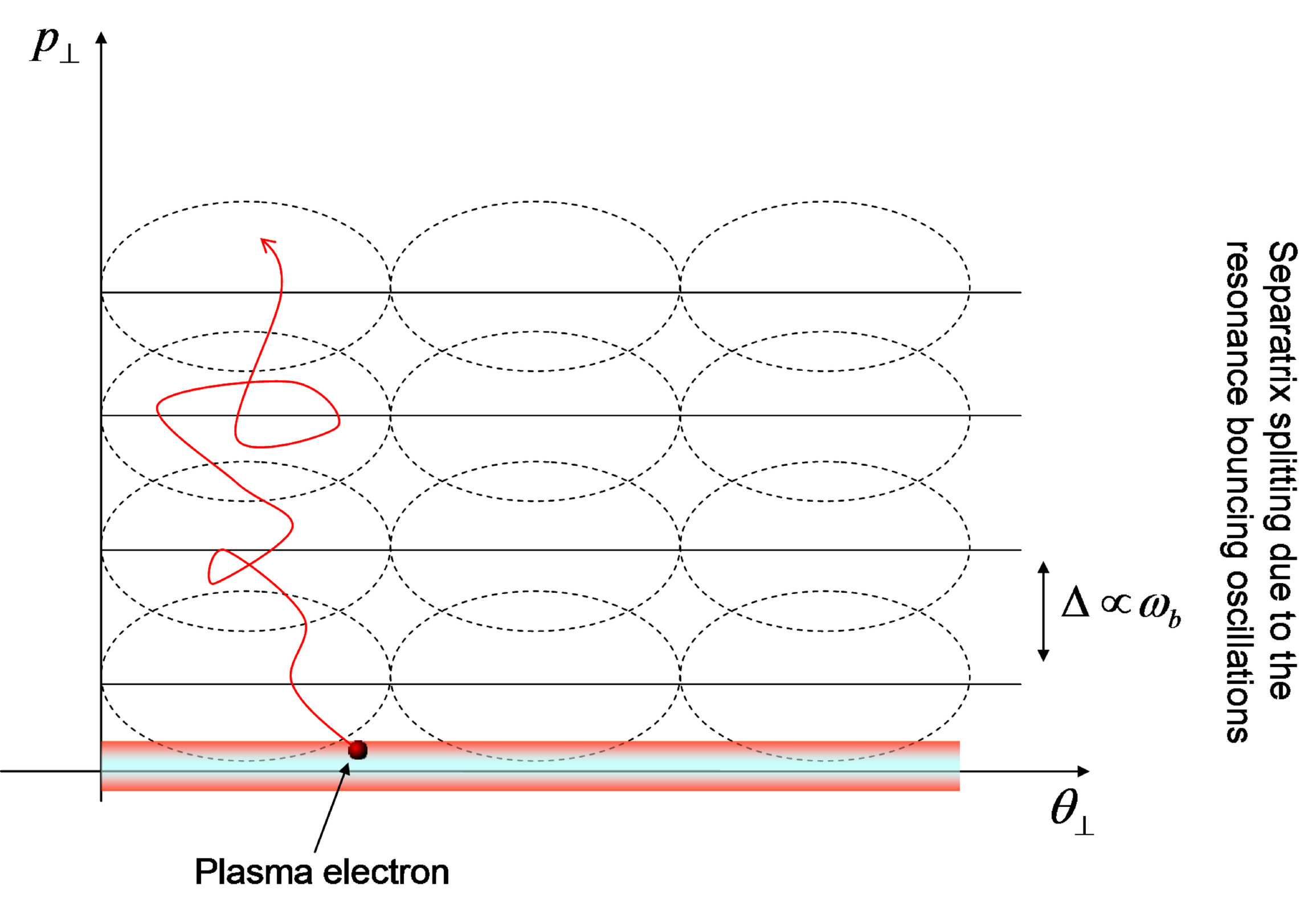}
\end{center}
\caption{\label{fig,separatrix_overlap} The many separatrices occurring in the proximity of the ECR zones: they are overlapped}
\end{figure}

It is clear that all the separatrices are overlapped, the distance between them being proportional to $\omega_{\rm b}$. The lowest separatrix intercepts the plasma slab; this means that electrons of the plasma can be trapped into the lowest separatrix. According to its own kinetic energy, to the phase with respect to the wave, and so on, the motion will evolve into the several separatrices in a complicated manner, taking into account also the nonlinearity in the wave--particle interactions. As the separatrices are overlapped, the electron perpendicular momentum (thus its energy) will increase more and more on passing from one separatrix to the subsequent one, as schematically shown in Fig.~\ref{fig,separatrix_overlap}. In the real situation the horizontal separatrix extension at the resonance is limited, and thus the electron will escape after a fixed time from the interaction zone. But because of its bouncing motion, it will pass again through the resonance. The former interaction was regulated by a stochastic motion in the phase space, so it will arrive again at the ECR with a completely random phase, thus being able to gain more energy. This process goes on for many passages through the resonance. However, the higher the electron energy, the higher is $\omega_{\rm b}$. When the bouncing frequency becomes comparable to the Larmor frequency, the separatrix splits and the electron cannot pass through them. This effect corresponds to the activation of some adiabatic invariants: the electron heating stops because the phase randomization process, strictly connected to the separatrix overlapping, \ced{no longer works}.

A more quantitative treatment of the heating process starts from the Canobbio parameters. Liebermann and Lichtenberg formulated the theory to calculate the maximum value of the electron energy for particles bouncing between the mirror, then passing many times through the resonance. Let $t_\rme$ be the time effectively spent by the electron in the ECR region:
\begin{equation}
\label{eq,tempoeffettivoECR}
t_\rme=\frac{0.71}{\omega}\left(\frac{2\omega}{\delta
v_{\perp{i}}}\right)^{{2}/{3}},
\end{equation}
where the perpendicular velocity $v_{\perp}$ is given by
\begin{equation}
v_{\perp}\propto\frac{e}{m}Et_\rme
\end{equation}
that is, in terms of $\delta$,
\begin{equation}
v_{\perp}\propto\frac{E}{\delta^{{2}/{3}}}
\end{equation}
and also
\begin{equation}
v_{\perp}\propto\frac{\sqrt{P}}{\delta^{{2}/{3}}}.
\end{equation}
By virtue of Eq.~\eqref{eq,tempoeffettivoECR}, we have that $t_\rme$
does not depend on $v_{\parallel{R}}$,  and it is inversely proportional to $v_{\perp{R}}$, i.e.\ the higher the transverse velocity, the less effective will be the particle acceleration.

For multiple passages through the ECR region, the
$v_{\perp}$ evolves according to
\begin{equation}
v_{\perp}^2=v_{\perp{0}}^2+v^2+2(vv_{\perp{0}}^2\cos\theta),
\end{equation}
where $v$ is the maximum velocity gained for each passage,
$\theta$ is a function of the phase and $vv_{\perp{0}}\cos\theta$
is the stochastic term.
After some passages, the mechanism explained above begins to play a role, the separatrix splits and the heating stops. This mechanism can be explained also in a different manner. As the velocity increases, some adiabatic invariants arise. Such invariants depend also on the wave electric field ($\vec{E}$). Defining by $W_{\rm S}$ the maximum value of the energy that can obtained through stochastic heating, we have
\begin{equation}
\label{eq,Ws}
W_{\rm S}=0.34eEL\sqrt{\frac{\bar{t}}{\tau_{\rm S}}},
\end{equation}
with
\begin{equation}
\label{eq,shift2pitime}
\tau_{\rm S}\sim4\pi\frac{m}{e}\frac{1}{B}\frac{L^2}{l^2},
\end{equation}
which is the time necessary to have a phase shift of $2\pi$, $L$ is the characteristic length of the mirror field, as can be seen by the $\vec{B}$ equation
\begin{equation}
\label{eq,Bz}
B_z=B_0\left(1+\frac{x^2}{L^2}\right),
\end{equation}
and
\begin{equation}
\label{eq,tsegnato}
\bar{t}=\sqrt{\left(\frac{mL}{eE}\right)}\left(1+\frac{l^2}{L^2}\right)^{{5}/{4}},
\end{equation}
corresponding to the resonance position $x=l$. In these conditions the electron is mirrored just inside the ECR zone \cite{LichtembergI,LichtembergII}.

The maximum energy achievable by the electrons will be
\begin{equation}
\label{eq,absstochbarrier}
W_{\rm B}\sim5W_{\rm S},
\end{equation}
and $W_{\rm B}$ is usually called the \textit{absolute stochastic barrier}.
To give an estimate of such a barrier, we can substitute some realistic values into the equations \eqref{eq,Ws}--\eqref{eq,absstochbarrier}, thus obtaining
\begin{equation}
W_{\rm S}\sim2E^{{3}/{4}}\ (\mathrm{V{\u}m^{-1}}) \propto P_{\rm RF}^{{3}/{8}},
\end{equation}
with $E$ and $P_{RF}$ the wave's electric field and power. As for ECRIS we have
\begin{equation}
0.1\leq E\leq10\ (\mathrm{kV{\u}cm^{-1}}).
\end{equation}
The absolute stochastic barrier will lie in the range
\begin{equation}
10{\u}\mathrm{keV} \leq W_{\rm B} \leq 270{\u}\mathrm{keV}.
\end{equation}

The estimation of the maximum energy achievable by ECRIS plasma electrons is realistic if compared with experiments. This is a remarkable result, as the single-particle theory may be considered somewhat oversimplified. However, such calculations \ced{fail} for modern ECRIS, like VENUS. Other possible heating mechanisms may be investigated thanks to numerical simulations in order to understand if the adiabatic invariants, activated at high electron energy, can be broken by additional effects not considered in the Liebermann and Lichtenberg theory.

Strictly speaking, it can be noted, by the separatrix treatment, that the heating may continue if the separatrix splitting does not occur, i.e.\ if additional perturbations to the particle motion change the separatrix amplitudes or provide an additional separatrix, in order to have a separatrix overlap again. In this way the phase randomization should continue even above the stochastic barrier.

The so-called `ECRIS standard model', which has traced the road for the development of electron cyclotron resonance ion sources in the past 20 years, is based on experimental evidence and mixes Geller's scaling laws and the high-B mode concept.
This model has permitted an average increase of about one order of magnitude per decade in the performance of ECR ion sources since the time of the pioneering experiment of R. Geller at CEA-Grenoble \cite{gr}. However, nowadays such a trend is limited not only by the technological limits of the magnets, but also by the experimental evidence of hot electron population formation, which becomes more important for high plasma density (i.e.\ for an operational frequency of 28{\u}GHz and higher) and for microwave power above 3{\u}kW.

The simplest picture to determine a relationship between the microwave frequency and the maximum achievable plasma density considers the $O$-\textit{mode cutoff frequency}, so that, writing it in terms of maximum density as a function of the field frequency, we obtain
\begin{equation}
\label{eq,cutoffSL}
n_{\rm cutoff}=\epsilon_0 m_\rme \frac{\omega^2}{e^2}.
\end{equation}
More generally, all the possible electromagnetic modes propagating in plasmas ($R,\ L,\ O,\ X$) suffer a cutoff. Then the maximum density of the plasma where these modes propagate cannot \ced{exceed} some fixed values. Only electrostatic modes are able to propagate in plasmas of whatever density (e.g.\ Bernstein modes).

The main consequence of Eq.~\eqref{eq,cutoffSL} is that, by increasing the electromagnetic field, the electron density can be increased. In magnetized plasmas the relationship between density and frequency may change, but the dependence of the extracted current on the square of the microwave frequency has been demonstrated in many experiments.
On the basis of these considerations and of the experimental results, 25 years ago  Geller proposed the following scaling laws \cite{g}, which for a long time have been the guideline for the ECRIS community:
\begin{eqnarray}
\label{eq,gellerlaws}
q_{\rm opt} & \propto & \log B^{{3}/{2}} , \\
I^{q+} & \propto & f^2 M_\rmi^{-1} ,
\end{eqnarray}
where $q_{\rm opt}$ is the optimum charge state, $B$ is the peak field of the magnetic trap, $f$ is the microwave frequency, $I^{q+}$ is the intensity of the charge state $q$ and $M_\rmi$ is the mass of ions.

These formulas have been verified in several ECRIS spread over the world.
However, although the microwave frequency was continuously increased, in order to fulfil Eq.~\eqref{eq,gellerlaws}, the increasing trend of the source performance \ced{began} to saturate, especially when it was clear that the MINIMAFIOS source \cite{mini} produced at 16.6{\u}GHz a mean charge state lower than the SC-ECRIS operating at 6.4{\u}GHz at MSU.\footnote{The MINIMAFIOS source was able to extract a CSD with $\langle q\rangle=9$, whereas the SC-ECRIS produced $\langle q\rangle=12$.}
It was clear that the confinement, which for the SC-ECRIS was much better than for MINIMAFIOS, was the other key parameter to ensure the exploitation of the benefits provided by the increase of the microwave frequency increase.

\section{Advanced research for ECRIS development}

The best compromise in terms of intense currents and charge states is obtained by ECRIS. Up to now the `ECRIS standard model' has traced the road for the development of ECRIS. Since its formulation (completed in 1990, when the HBM concept was proposed), this model has been proven by many experiments carried out with different sources, operating at different ECR frequencies.
The main rules were confirmed by experiments performed at MSU-NSCL in 1993--94 and in 1995.
At the powers, magnetic fields and microwave frequencies used up to now, the ECRIS obey the standard model:
the extracted current increases strongly as the microwave frequency increases, \ced{but only the increase of the magnetic mirror ratio ensures that the increased plasma energy content can be efficiently exploited, by making more stable the confinement.}
Assuming the conclusions of the high-B mode concept, and linking them with the experimental results, it is known that the magnetic field distribution in the plasma chamber must obey the following rules:
\begin{enumerate}
\item [(a)]  the radial magnetic field value at the plasma chamber wall must be $B_{\rm rad} \geq 2 B_{\rm ECR}$;
\item [(b)]  the axial magnetic field value at injection must be $B_{\rm inj} \simeq 3B_{\rm ECR}$ or more;
\item [(c)]  the axial magnetic field value at extraction must be about $B_{\rm ext} \simeq B_{\rm rad}$; and
\item [(d)]  the axial magnetic field value at minimum must be in the range $0.30 < {B_{\rm min}}/{B_{\rm rad}} < 0.45$.
\end{enumerate}
Then, according to the standard model, the ECRIS development is strictly linked to the improvements of superconducting magnets and of the microwave generator technology.
\ced{There is no evidence for scaling laws for} the microwave power and the ECR heating process (ECRH), because the relation between the power and the magnetic field is not so simple. Some authors studied the RF coupling to the plasma in terms of the maximum power rate per unit volume and its relationship with the beam intensity produced by different ECR ion sources \cite{hitz}. Only recently it was observed that the efficacy of the RF to plasma energy transfer depends also on the amount of wave energy coupled to the plasma chamber. Thus the cavity design, and in particular the microwave injection geometry, is of primary importance for a high RF energy transmission coefficient into the plasma chamber.

Note that the problem of the wave energy transmission into the plasma must be divided into two parts: the first is connected to the microwave generator--waveguide--plasma chamber coupling, while the second one \ced{relates to} the wave--plasma interaction.
The standard model does not allow one to argue for any prediction about the mechanism at the basis of electron and ion dynamics for different magnetic field profiles and for different microwave frequencies. Some preliminary experimental results obtained with a third-generation ECRIS (VENUS) \ced{showed the limitations} of the standard model, in particular with regard to the magnetic field scaling. Although the standard model has permitted strong improvement in the ECRIS performance up to the third-generation ones (ECRIS that operate at 28{\u}GHz and with magnetic field up to 4{\u}T), we expect that the technological limits can be overcome by means of a better understanding of plasma physics, in particular of the ECR heating mechanism.

Since 1994 (first evidence for the two-frequency heating effect), several possible ways to improve ECRIS performance overcoming the \ced{`brute force'} scaling of the main operative parameters have been investigated worldwide.

\subsection{Two-frequency heating}

Historically, this can be considered the first non-conventional method of plasma heating outside the scheme traced by the standard model.
Since 1994, the so-called two-frequency heating (TFH) has been used \cite{xie,scott}
to improve the HCI production by feeding the plasma with
two electromagnetic waves at different frequencies instead of
one. In some cases, even three or more close frequencies or
a white noise generator (WNG) \cite{kaw} have been used. The TFH has been
demonstrated to be a powerful method. For example, in the case
for $^{238}$U \cite{xie}, it increased the production of higher
charge states (from $35+$ to $39+$) by a factor of 2--4
and shifted the peak charge state from $33+$ to $36+$. Unfortunately,
neither the relationship between the two frequencies
nor the respective power was unequivocally determined. In fact,
any source features a different set of parameters and the
optimization is done empirically, just by looking to the maximization
of beam current.

Several qualitative explanations have
been given about this phenomenon, all related to the increase
of the average electron temperature $T_\rme$ and the ionization rate
by assuming that the crossing of two resonance surfaces helps
the electron to gain more energy. This simple picture does not
explain the reason for the relevant changes in the charge state
distribution (CSD) for different pairs of frequencies (even for
the case of minor changes, let us say a few MHz over $14$ or
$18${\u}GHz), which can be explained by the frequency tuning effect. It is important to underline that, even in the case of TFH
applied to many existing sources, a TWT is often used, the other
being a klystron-based generator. The choice of TWT allows
experimentalists to vary the second frequency slightly. As an example, a strong variation of ECRIS performance has been observed in TFH operations \cite{scott}. It
was observed that the production of O$^{7+}$ with 60{\u}W emitted
by the TWT at the optimum frequency gives the same effect
as 300{\u}W from the fixed-frequency klystron. The maximum
current is obtained by means of klystron at 427{\u}W and TWT at
62{\u}W ($I = 66${\u}e{\u}$\rmmu$A). In order to obtain the same current, they
needed $P_{\rm RF} = 800${\u}W by the klystron in SFH. Furthermore, in
the case of TFH, the current increases almost $20\%$ (from 57
to 66{\u}e{\u}$\rmmu$A) when the TWT emitted frequency shifts from 11.06
to 10.85{\u}GHz, the klystron and TWT emitted power \ced{remaining constant}.
Thus, the TFH is an effective method to increase the extracted
current from ECRIS, but it can be fully exploited only
by means of frequency tuning. Several measurements have been carried
out with the SERSE ion source. The TFH has been used
for operation at either 14 or 18{\u}GHz, with a clear advantage with
respect to SFH. It was not successful when a 28{\u}GHz gyrotron and an 18{\u}GHz klystron were used at the same time \cite{gir}, and
that is evidence that fixed-frequency generators are not adequate for TFH. The performance obtained by a two-fixed-frequency klystron
was also not as good as that obtained by a fixed-frequency generator plus a TWT amplifier.

As an example, for tin production, a current of the order of 2--3{\u}e{\u}$\rmmu$A
was measured for the charge state $29^+$ when a power
of 1.4{\u}kW from the 18{\u}GHz klystron and a power of 1.0{\u}kW
from the 14.5{\u}GHz klystron were used. By keeping
all the parameters and the 18{\u}GHz klystron power unchanged, a TWT amplifier
operating in the range of 8--18{\u}GHz replaced the 14.5{\u}GHz
klystron. Even if the maximum power from the TWT was only
200{\u}W, it permitted one to obtain more than 3{\u}e{\u}$\rmmu$A of the charge
state $29^+$ with a frequency of 17.600{\u}GHz. Another interesting
operating point was observed at the frequency around
17.000{\u}GHz, which was less critical than any other set-up and
permitted one to obtain the state $30^+$ with the same beam
current, by means of only minor changes of the gas pressure. It
must be underlined that the vacuum inside the plasma chamber
improved by decreasing the total power from 2.4 to 1.6{\u}kW,
which permitted one to move the CSD to higher charge states
and to have an excellent reproducibility, even after \ced{some} weeks. In order to compare these results with the
performance in SFH mode, the Sn$^{30+}$ current obtained with
the same power from the 18{\u}GHz klystron was only 0.7{\u}e{\u}$\rmmu$A. All these data are explained by the description of the microwave--plasma interaction given here and they demonstrate the advantage of TFH operations, for highly charged ion production
when the chosen frequency has the right electromagnetic field
distribution in the plasma chamber.
Moreover, the closer frequencies are the best choice.

To explain TFH on the basis of the fundamental processes occurring during the electron--wave interaction, numerical simulations based on a pure Monte Carlo approach were developed.
To be adequately modelled, the TFH effects can be depicted in the following way:
\begin{enumerate}
\item We double the resonance zone width, so that electrons that do not gain energy during the first crossing may have another possibility to be heated by the second one. Figure \ref{fig,TFH_resonance_zones} shows the locations of these two resonance regions inside the plasma chamber calculated for the initial set-up of our simulations.
\item Preliminary simulations have shown a significant role played by the ECR interaction in recovering electrons otherwise contained in the magnetic loss cones. This effect (a deterministic plug-in of cold electrons) is currently deemed to play the main role in confining most of the plasma inside the volume embedded by the resonance surface. Adding a second resonance, electrons that have not been plugged by the first wave can be recovered by means of the second one, with a further increase of the plasma confinement.
\end{enumerate}

A series of simulations has been carried out in order to verify
the effect of the TFH. In the case in Fig.~\ref{fig,TFH_resonance_zones}, the two resonance zones are far \ced{apart}. This allows the electrons to randomize their
gyro-phase with respect to the electromagnetic wave. The numerical study has dealt with the case of two electromagnetic
waves at different frequencies supplying the mode
TE$_{4,6,22}$ (1000{\u}W) and the mode TE$_{4,4,23}$ (300{\u}W), respectively.
It is important to note in Table \ref{tab,tab_Davide_Lanaia} that, by means of the
TFH, the final energy is almost \ced{twice as high as} in the
case of SFH at 18{\u}GHz, whereas the amount of recovered
electrons (i.e.\ equal to $38.2\%$) is \ced{twice that} obtained
in single-frequency operation (18{\u}GHz; 1000{\u}W).

\begin{figure}[htbp]
\begin{center}
\includegraphics[width=0.5\textwidth]{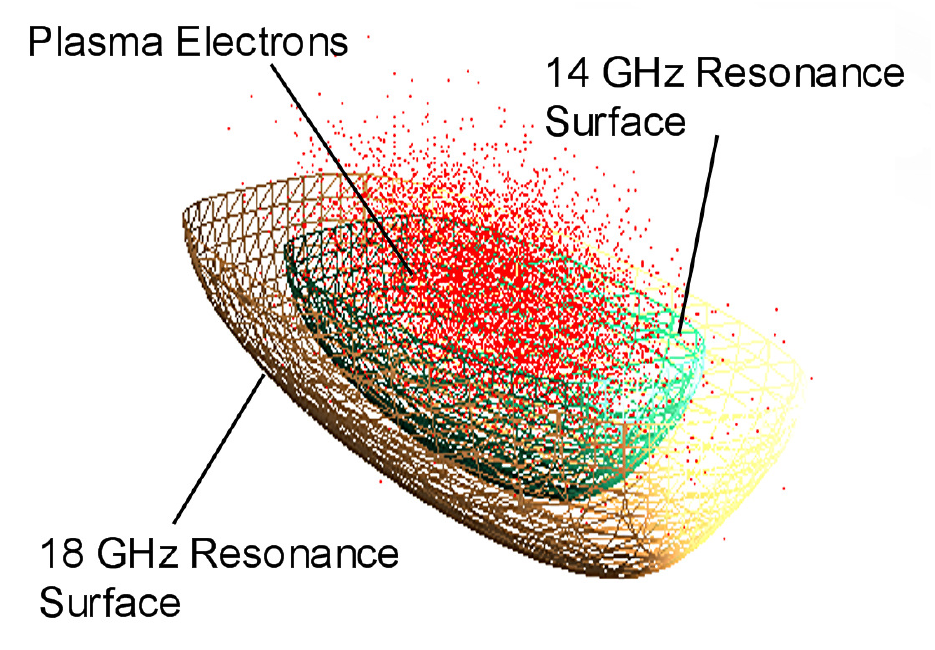}
\end{center}
\caption{\label{fig,TFH_resonance_zones} Section of the two resonance surfaces showing the position of the plasma electrons}
\end{figure}

\begin{table}[htbp]
\caption{\label{tab,tab_Davide_Lanaia} Percentage of recovered electrons (i.e.\ electrons that are recovered by the electromagnetic field) and the electron
energy after 50{\u}ns for different heating configurations (SFH and TFH) and at different powers.}
\begin{center}
\begin{tabular}{ccccc}
\hline\hline
\textbf{Frequency (GHz)} & \textbf{Power (W)} & \textbf{Recovered electrons (\%)} & \textbf{Energy (keV)} \\
\hline
14 + 18 & 1000 + 300    & 38.2  & 1.9589 \\
  18      & 2000        & 35.2  & 1.2612 \\
  18      & 1000        & 19.2  & 1.0141 \\
  14      & 300         & 12.9  & 1.5509 \\
\hline\hline
\end{tabular}
\end{center}
\end{table}

\subsection{Frequency tuning effect}

Many experiments over the past years have shown that significant improvements of ECRIS performances (in terms of both total extracted current and highly charged ion production) are obtainable by slightly varying the microwave frequency in the case of single frequency heating (SFH), defined as a `frequency tuning' effect.
It was known that the variation of the frequency increases the electron density, \ced{and thereby} improves the ECRIS performance. However, this occurs only for variations of the order of GHz, and it is strictly connected with the increase of the cutoff density.

On the contrary, several experiments have demonstrated that even slight variations of the pumping wave frequency may lead to strong variation of the extracted current for SFH. Such variations are of the order of MHz to GHz, but they are sufficient to strongly change the wave propagation inside the plasma chamber and the mode pattern on the resonance surface.

The first evidence that significant improvements can be obtained by varying the frequency of the microwaves was given by the different performance observed for the SERSE and CAESAR ion sources when fed by a klystron-based or a TWT-based generator \cite{celona,gammino,von} at either 14 or 18{\u}GHz.
Similar results came from experiments performed at ORNL and at JYFL \cite{kaw}.

\begin{figure}[htbp]
\begin{center}
\includegraphics[width=0.75\textwidth]{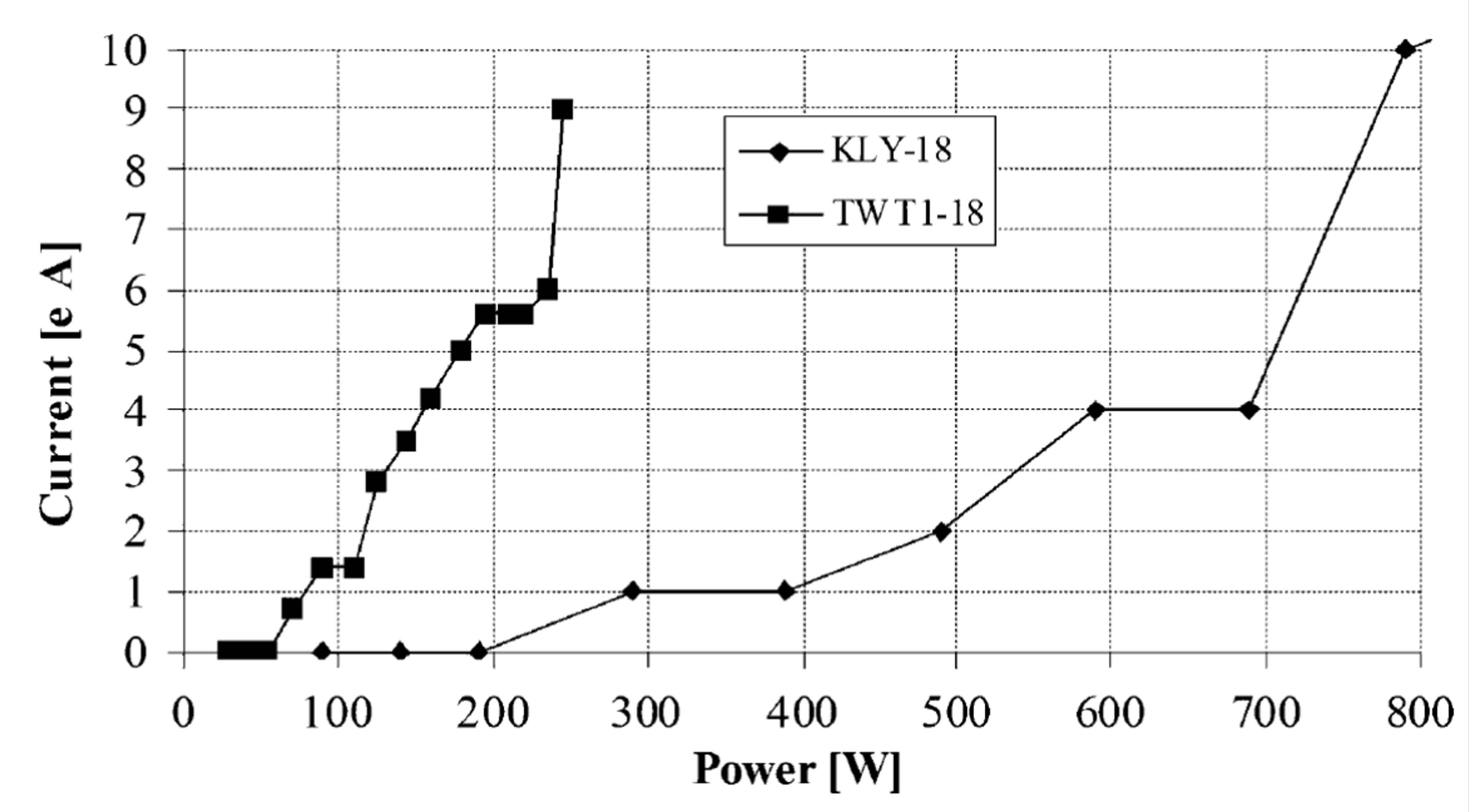}
\end{center}
\caption{\label{fig,TWTvsKLY} Comparison between trends of O$^{8+}$ at 18{\u}GHz for a klystron (up to 800{\u}W) and a TWT}
\end{figure}

Figure \ref{fig,TWTvsKLY} shows that a remarkable increase in the production of O$^{8+}$ can be obtained by using the TWT instead of klystrons. In particular, the current of O$^{8+}$ obtained with the TWT at 240{\u}W is obtained by the klystron at 800{\u}W, i.e.\ a power three times higher. It is important to underline that the two generators operated at two different but close frequencies. In particular, the klystron was operating at 18.0{\u}GHz, while the TWT was at 17.9{\u}GHz (both the amplifiers were fed by \ced{dielectric resonator-type oscillators}). Such difference of performance was initially explained as a greater frequency dispersion of the TWT, but a series of measurements carried out with the two microwave generators, by means of a spectrum analyser, pointed out that the spectrum of the emitted radiation of a TWT is similar to that of the klystron. So the only difference between the two generators is the output frequency, with the further possibility for the TWT to vary the emitted frequency, thus optimizing the source performance.

The experimental set-ups used for the confirmation of the frequency tuning effect were the SUPERNANOGAN ion source of CNAO, Pavia, and the CAPRICE source at GSI, Darmstadt, \ced{which demonstrated that the frequency tuning strongly affects also the ion beam formation.}
Figure \ref{fig,Supernanogan_current} shows the current for C$^{4+}$ obtained with the SUPERNANOGAN ion source in 2005 versus the microwave frequency, keeping the power and all the other source parameters unchanged. The figure features strong fluctuations of the extracted current in a frequency span of about 90{\u}MHz.

\begin{figure}[htbp]
\begin{center}
\includegraphics[width=0.50\textwidth]{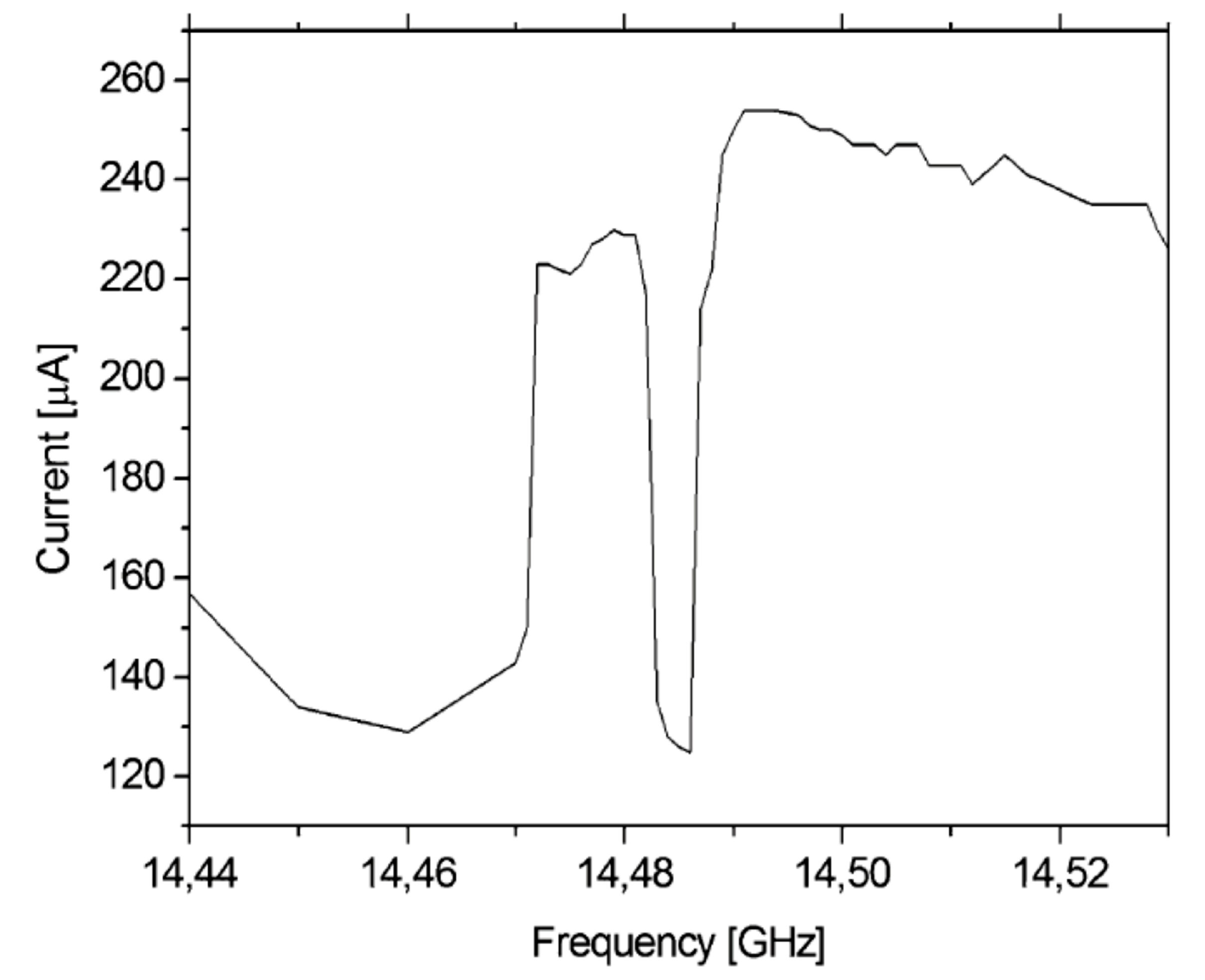}
\end{center}
\caption{\label{fig,Supernanogan_current} Trend of the analysed C$^{4+}$ current for the SUPERNANOGAN ECR ion source versus the RF frequency}
\end{figure}

The frequency was changed in the range 14.44 to 14.53{\u}GHz with a step of 1{\u}MHz and it was observed that changes of a few MHz changed the C$^{4+}$ current even by $70\%$.
Further experiments including the final validation of the frequency tuning effect were carried out with the SUPERNANOGAN source of CNAO. It permitted an increase by $30\%$ in its performance for C$^{4+}$ and by $50\%$ for H$^{3+}$ (these ions are particularly requested for medical applications), and an additional increase of reliability and availability figures was registered.

Interesting results come from the GSI experiment carried out in 2007. On that occasion, as stated before, the frequency tuning was demonstrated to strongly affect also the beam shape, along with the extracted current for each charge state.

As an example, Fig.~\ref{fig,beamshapevariation} reports the shape evolution of a helium beam recorded on a viewer.
The two helium charge states can be observed as well as the aberrations introduced by
the hexapole. Furthermore, Fig.~\ref{fig,beamshapevariation} evidently shows that the beam intensity distribution
is inhomogeneous and that this distribution changes with the microwave frequency,
while keeping all the other parameters constant.

\begin{figure}[htbp]
\begin{center}
\includegraphics[width=0.70\textwidth]{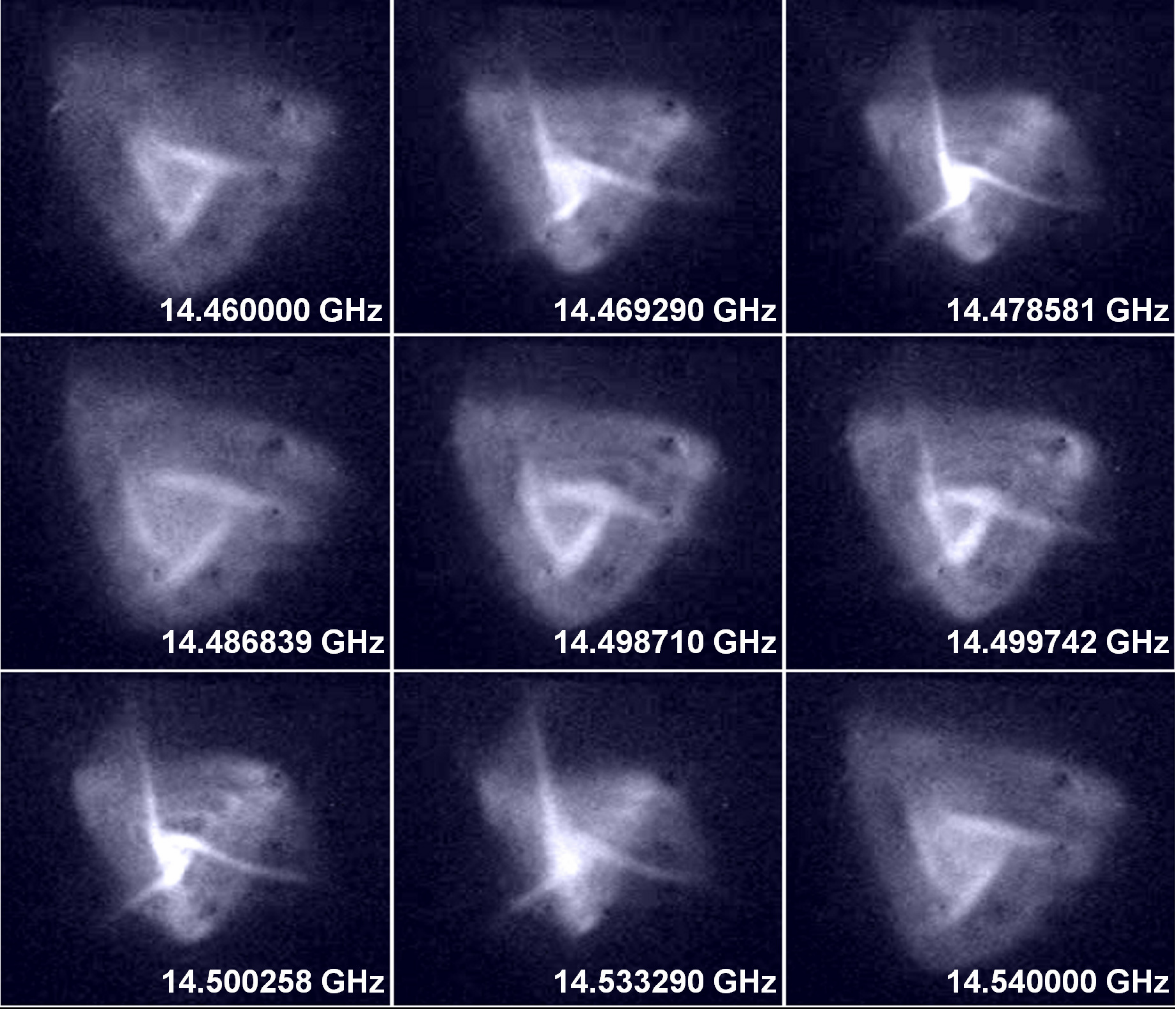}
\end{center}
\caption{\label{fig,beamshapevariation} Variation of the extracted beam shape for different microwave frequencies}
\end{figure}

Measurements of the $S_{11}$ scattering parameter were carried out in parallel. The $S_{11}$ parameter is connected with the reflection coefficient, and the measurements have shown that, for some frequencies, the amount of reflected power increases, thus demonstrating that some resonances are present inside the cavity with or without plasma. In addition, fluctuation in the extracted power occurs also in the case of quasi-constant reflection coefficient.

A series of simulations were carried out in order to verify the effect of different microwave frequencies on the electron dynamics.
The mode pattern over the resonance surface was investigated.

The resonance surface can be calculated numerically once the magnetic field structure is known, as it corresponds to that surface with constant field equal to
${m \omega_{\rm RF}}/{q}$. Figure \ref{fig,EMfield_over_the_resonance_surface} features the resonance surface of the SERSE source calculated numerically by means of MATLAB. Note that, by varying the frequency by about 50{\u}MHz, the field distribution over the surface changes remarkably. In particular, if one consider the zones of maxima, the electric field is more than $10^2$ higher in the case of Fig.~\ref{fig,EMfield_over_the_resonance_surface}(b) than in the case of Fig.~\ref{fig,EMfield_over_the_resonance_surface}(a).

\begin{figure}[htbp]
\begin{center}
\includegraphics[width=0.90\textwidth]{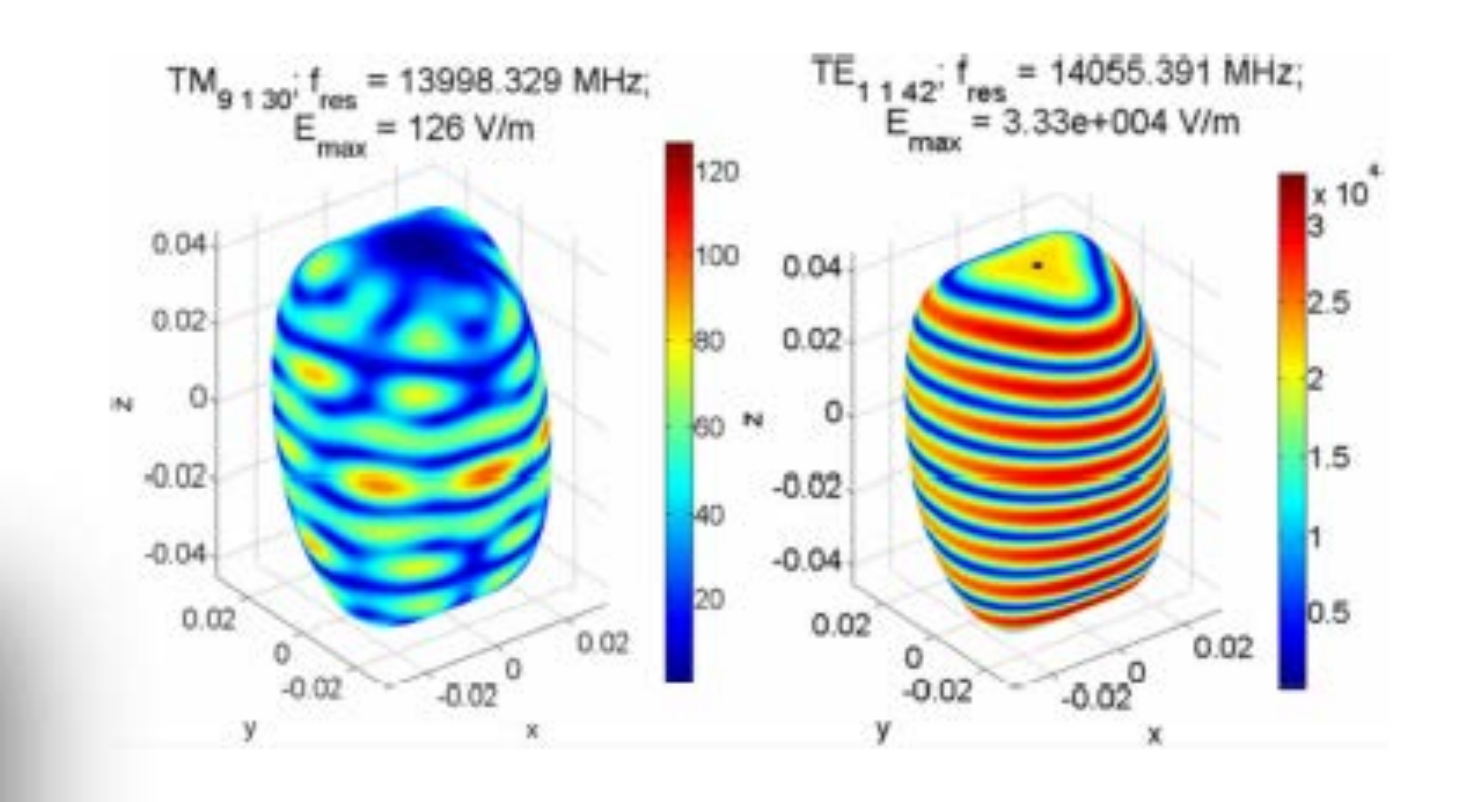}
\end{center}
\caption{\label{fig,EMfield_over_the_resonance_surface} Simulated resonance surface of the SERSE source with the distribution of the electromagnetic field (false colour representation).}
\end{figure}

Thus the excitation of near modes, even with distances of a few MHz, strongly changes not only the maximum field over the surface, but also the distribution of the zones of minima and maxima.

Simulations carried out considering different modes have demonstrated that some of them are able to heat the electrons up to energies much higher than others (even ten times higher). Figure \ref{fig,simulated_frequency_tuning} features the strong fluctuations of the energy gained by the electrons after 50{\u}ns. The simulations have been done in the proximity of 14{\u}GHz, and for the SERSE source parameters (in terms of magnetic field structure, plasma chamber dimensions and input RF power). The maximum energy after 50{\u}ns has been reached when the TE$_{1,1,42}$ mode has been used for plasma heating, whereas the minimum energy has been reached for the mode TE$_{9,1,30}$. Note that these two extreme cases correspond to the field distribution shown in Fig.~\ref{fig,EMfield_over_the_resonance_surface}. In our study the net power flowing into the chamber was kept constant, and therefore the observed effects are only due to the different field patterns. In fact, it is clear from the figure that the TE$_{1,1,42}$ has a higher electric field over a greater part of the resonance surface with respect to the TE$_{9,1,30}$ mode.

\begin{figure}[htbp]
\begin{center}
\includegraphics[width=0.60\textwidth]{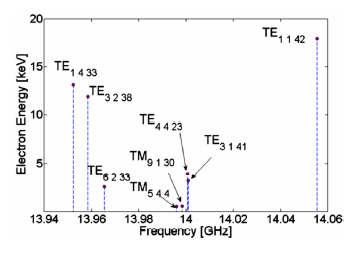}
\end{center}
\caption{\label{fig,simulated_frequency_tuning} Simulation of electron heating rapidity for different modes}
\end{figure}

In order to explain the simulation results shown in Fig.~\ref{fig,simulated_frequency_tuning}, we have to take into account that the \ced{greatest} amount of energy transfer from waves to electrons occurs on the resonance surface because of the ECR condition. On the other hand, the electrons gyrate around the field lines with the Larmor frequency, with their guiding centre moving along the magnetic lines. Then the structure of the magnetic field determines the crossing points of the electron trajectories with the resonance surface.

Figure \ref{fig,EMfield_&_trajectory} points out that only in particular zones of the resonance surface does energy transfer occur, and the amount of energy gained by the electrons depends strongly on the mode structure, as only in the case of high-field zones in the proximity of the trajectory crossing points is it possible for the electron to gain very high energies.

\begin{figure}[htbp]
\begin{center}
\includegraphics[width=0.70\textwidth]{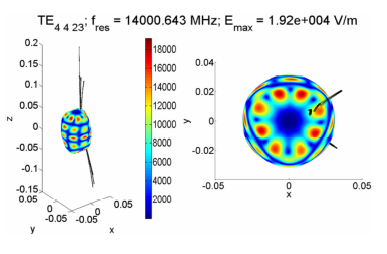}
\end{center}
\caption{\label{fig,EMfield_&_trajectory} The electron trajectory and the simulated resonance surface with the electromagnetic field. The crossing point between the electron trajectory and the resonance surface is of primary importance for the electron heating.}
\end{figure}

The situation here describes reasonably the CAPRICE results, and the simulation results demonstrate that variation of frequency of the order of tens of MHz (over a pumping frequency of 14{\u}GHz) strongly affect the electron heating, and thus also the source performance.

The results obtained with SERSE fed with TWT or klystron microwave generators can also be explained thanks to the simulation results presented above. The TWT-based generators, in fact, are able to change the output frequency, and so by using them it is possible to inject into the plasma chamber the best frequency in order to maximize the heating rapidity $v_{\rm h}$.

Note that the higher the heating rapidity, the higher is the replacement of the electrons lost from the confinement because of the Spitzer collisions. This effect is of primary importance because it permits one to obtain high densities of warm electrons.

The electromagnetic field patterns over the ECR surface for several frequencies are  shown in Fig.~\ref{fig,EMfield_over_the_resonance_surface}. It is clear that on passing from 14 to 18{\u}GHz the pattern of the resonant modes over the ECR surface changes considerably. This means that after the effects due to the scaling laws, we have to take into account the differences in the field patterns when the frequency changes.

\section{Highly charged ions -- perspectives}

There is no doubt among experts that the incredible rate of increase of ion source performance cannot be sustained in the future, but some of the recent results have demonstrated that a better comprehension of plasma physics may permit additional increases of average charge states and currents.
Further tests in \textit{ad~hoc} test benches are under way to improve the knowledge of the new plasma heating schemes, and the saturation of magnet performances may not be a limiting factor in the future. The proposal of fourth-generation ECRIS with operational frequency above 50{\u}GHz is appealing but it is questionable if a similar device may run with stable and reproducible beams, to the extent that heating is not controlled.

\section*{Acknowledgements}

\ced{The chapter presented here has been prepared in strict collaboration with David
Mascali and Giuseppe Torrisi, whose active contributions permitted my schematic lesson notes to be translated into a readable form.}

\newpage

\end{document}